\pdfoutput=1
\documentclass[prm,10pt,aps,longbibliography,superscriptaddress,nofootinbib,twocolumn]{revtex4-2}

\usepackage[dvipsnames]{xcolor}
\usepackage{amsmath,amssymb}
\usepackage{graphicx}
\usepackage{footmisc}
\usepackage{bbold,soul}
\usepackage{lipsum}
\usepackage{enumitem}
\usepackage{hyperref}
\usepackage{physics}
\usepackage{watermark}
\usepackage{array}
\usepackage{multirow}
\usepackage{tabularx}
\usepackage{listings}
\usepackage{lineno}
\usepackage{float}
\usepackage{multirow}
\usepackage{array}
\usepackage{makecell}
\usepackage{tcolorbox}
\usepackage{booktabs}
\usepackage{color}
\usepackage[utf8]{inputenc}
\usepackage{forest}
\usepackage{xcolor}
\usepackage{bm}

\hypersetup{
    colorlinks=true,
    linkcolor=red,
    filecolor=black,
    urlcolor=darkblue,
    citecolor=red,
}

\definecolor{codegreen}{rgb}{0,0.6,0}
\definecolor{codegray}{rgb}{0.5,0.5,0.5}
\definecolor{codepurple}{rgb}{0.58,0,0.82}
\definecolor{orange}{rgb}{1, 0.13, 0.2}
\definecolor{backcolour}{rgb}{0.97,0.97,0.97}
\definecolor{lightgreen}{rgb}{0.8, 1, 0.8}
\definecolor{lightblue}{rgb}{0.8, 0.9, 1}
\definecolor{lightyellow}{rgb}{1, 1, 0.8}
\definecolor{darkblue}{rgb}{0.0,0,0.75}
\definecolor{lightgray}{rgb}{0.925,0.925,0.925}
\lstdefinestyle{mystyle}{
    backgroundcolor=\color{backcolour},
    commentstyle=\color{codegreen},
    keywordstyle=\color{blue},
    numberstyle=\tiny\color{codegray},
    stringstyle=\color{orange},
    basicstyle=\footnotesize,
    breakatwhitespace=false,
    breaklines=true,
    captionpos=b,
    keepspaces=true,
    numbers=left,
    numbersep=5pt,
    showspaces=false,
    showstringspaces=false,
    showtabs=false,
    tabsize=2,
    basicstyle=\footnotesize\ttfamily
}

\newcommand{\vecmu}{\boldsymbol{\mu}}

\newcommand{\code}[1]{\texttt{#1}}

\newcommand\footnoteref[1]{\protected@xdef\@thefnmark{\ref{#1}}\@footnotemark}

\newcommand{\filler}[1]{\colorbox{lightblue}{\footnotesize\textsf{#1}}}
\newcommand{\fillerPara}[1]{%
    \noindent\begin{minipage}[t]{0.45\textwidth}%
        \colorbox{lightblue}{%
            \begin{minipage}[t]{\linewidth}%
                \footnotesize\sffamily#1%
            \end{minipage}%
        }%
    \end{minipage}%
}

\renewcommand{\fillerPara}[1]{}
\renewcommand{\filler}[1]{}

\newlength{\hypothesisindent}
\newcounter{hypothesis}

\setlist[description]{leftmargin=1.0em,labelindent=0.5em}

\begin{document}

\title{OmniXAS: A Universal Deep-Learning Framework for Materials X-ray Absorption Spectra}

\author{Shubha R. Kharel}
\email{skharel@bnl.gov}
\affiliation{Computing and Data Sciences Directorate, Brookhaven National Laboratory, Upton, New York 11973, USA}

\author{Fanchen Meng}
\affiliation{Center for Functional Nanomaterials, Brookhaven National Laboratory, Upton, New York 11973, USA}

\author{Xiaohui Qu}
\affiliation{Center for Functional Nanomaterials, Brookhaven National Laboratory, Upton, New York 11973, USA}

\author{Matthew R. Carbone}
\email{mcarbone@bnl.gov}
\affiliation{Computing and Data Sciences Directorate, Brookhaven National Laboratory, Upton, New York 11973, USA}

\author{Deyu Lu}
\email{dlu@bnl.gov}
\affiliation{Center for Functional Nanomaterials, Brookhaven National Laboratory, Upton, New York 11973, USA}

\date{\today}

\begin{abstract}

  X-ray absorption spectroscopy (XAS) is a powerful characterization technique
  for probing the local chemical environment of absorbing atoms.
  However, analyzing XAS data presents with significant challenges, often requiring
  extensive, computationally intensive simulations, as well as significant domain expertise.
  These limitations hinder the development of fast, robust XAS analysis
  pipelines that are essential in high-throughput studies and for autonomous
  experimentation.
  We address these challenges with OmniXAS, a framework that contains a suite of transfer learning
  approaches for XAS prediction, each uniquely contributing to improved
  accuracy and efficiency, as demonstrated on K-edge spectra database covering eight 3d
  transition metals (Ti -- Cu).
  The OmniXAS framework is built upon three distinct strategies.
  First, we use M3GNet [\emph{Nat. Comput. Sci.},  2, 718 (2022)]
  to derive latent representations of the local chemical environment of absorption
  sites as input for XAS prediction, achieving significant improvements
  over conventional featurization techniques.
  Second, we employ a hierarchical transfer learning strategy, training a
  universal multi-task model across elements before fine-tuning for
  element-specific predictions. Models based on this cascaded approach after element-wise fine-tuning
  outperform element-specific models by up to 69\%.
  Third, we implement cross-fidelity transfer learning, adapting a universal
  model to predict spectra generated by simulation of a different fidelity with a much higher computational cost. This approach improves prediction accuracy by up to
  11\% over models trained on the target fidelity alone.
  Our approach significantly boosts the throughput of XAS modeling by orders of magnitude as compared to first-principles simulations and is extendable to XAS prediction for a broader range of elements. The proposed transfer learning framework is generalizable to enhance deep-learning models that target other properties in materials research.
\end{abstract}

\maketitle

\section{Introduction}

X-ray absorption spectroscopy (XAS) is a widely used materials characterization
technique in a broad array of scientific research
fields~\cite{rehr2000theoretical,penner1999x,de2008core}, such as condensed
matter physics, materials science, chemistry and biology. XAS is element-specific and its
near-edge region, known as X-ray absorption near-edge structure (XANES),
contains rich information of the local chemical environment of the absorbing
site (e.g., oxidation state, coordination number and local symmetry).
Therefore, XANES measurements provide important insights into the structural
and electronic properties of the sample, which are needed for mechanistic
understanding of the underlying physical and chemical processes.

However, XANES analysis is non-trivial, as the spectral function is a complex
representation of the underlying atomic structure and electronic structure associated with the short-range order.
Traditionally, extracting the information
from XANES spectral features heavily relies on empirical
fingerprints~\cite{farges1997ti} from well-established experimental standards and/or
first-principles simulations. These traditional XANES analysis
approaches require strong domain expertise and, when combined with
first-principles simulations, can be computationally demanding, creating a practical barrier for many researchers.

On the other hand, rapid advances in synchrotron X-ray facilities enables
XAS measurements with high time and energy resolution. For example, using a quick XAS scanning method,
a spectrum can be measured with 10 ms time resolution~\cite{muller2016quick}; the energy resolution in
high-energy-resolution fluorescence-detected (HERFD) XAS can reach below the
core-hole lifetime broadening~\cite{de2002spectral}. 
Such experimental instrumental development creates the demand of XANES analysis of large scale temporal data.
In addition, the deployment of autonomous
experimentation pipelines~\cite{yager2023autonomous,maffettone2023self,carbone2024flexible} makes a strong case for
real-time data analysis. To address these emerging challenges, a robust
data-driven XANES analysis approach is needed to lower the barrier to entry for
non-experts, reduce computational cost and accelerate throughput. In
practice, a data-driven XANES pipeline requires multiple key building blocks,
including (a) workflow software for high-throughput XANES simulations, (b) large and diverse
simulated XANES spectra databases, and (c) tailored machine learning (ML) models
that capture the structure-spectrum relationship.

In the past several years, significant progress has been made in this field.
Systematic multi-code XANES benchmarks~\cite{Meng2024} were carried out to
quantify the effects of key approximations and implementations in simulation,
as well as to determine the converged parameters for spectral simulations.
Workflow software~\cite{newville2013larch,kas2021advanced,carbone2023lightshow} has also been developed to
standardize the spectral input file generation and ensure data reproducibility.
Concurrently, the corpus of open-access, publicly available simulated XANES spectra
databases~\cite{mathew2018high,torrisi2020random,chen2021database,guda2021understanding,rankine2022accurate,guo2023simulated,cao2024atomic}
continue to grow, thus facilitating the development of ML models for XANES
analysis.

ML models trained to predict spectra from atomistic structure, i.e. spectroscopy surrogate models, are of particular importance, since once trained, they can bypass expensive first-principles simulations. In the same spirit of neutral network potentials
(NNPs)~\cite{behler2007generalized}, which are used to predict energies and forces
from atomistic geometries, ML
models can be trained on first-principles data to predict XANES spectra (or
more broadly, excited state properties, e.g. spectral functions) with the
accuracy of \emph{ab initio} methods but at a fraction of the computational
cost. Spectroscopy ML surrogates therefore make it feasible to tackle complex materials with
up to thousands atoms in a supercell.
Several ML models have been developed for XANES, such as graph neural network
(GNN) and multilayer perceptron (MLP) models for C, N, and O K-edge
XANES~\cite{carbone2020machine,ghose2023uncertainty} of the QM9 molecular
database~\cite{ramakrishnan2014quantum}, C K-edge XANES of amorphous
carbon~\cite{kwon2023spectroscopy,kwon2023harnessing}, K-edge XANES of Fe
compounds~\cite{rankine2020deep} and 3d transition metal
complexes~\cite{rankine2022accurate}. As these models represent a limited scope in materials and chemical space
(e.g., finite systems and a small sub-space of the extended systems), there is
still an unmet need to develop~\textit{general} ML models to predict XANES in a broad
materials space.

A fast and accurate spectral prediction surrogate model can play an important role in the XAS analysis pipeline. For a given target measured XAS spectrum, one can use an efficient search algorithm to sample the relevant material space and identify candidate structures. The corresponding XAS spectra are compared with experiment to identify the best match. For complex material systems (e.g., amorphous materials, defects, nanostructures, liquid systems and interfaces), the material space can be very large and each candidate structure may contain hundreds or thousands of atoms. It becomes impractical to perform a huge number of first-principles XAS simulations on very large systems in a high-throughput manner. A high-fidelity XAS surrogate model is the solution to this problem. The significance of this approach was demonstrated in the recent study of amorphous carbon~\cite{kwon2023spectroscopy}, where a diffusion model for structure sampling was combined with an XAS surrogate model to analyze the measured C K-edge XANES spectra.

In this work, we address the need for a generalizable ML
framework to predict XANES spectra from atomic structures.
Our approach leverages transfer learning and domain adaptation techniques to develop
universal models. By adopting M3GNet-derived latent features~\cite{chen2022universal} to predict K-edge XANES spectra for a range of 3d transition metals, we demonstrate that this method not only
outperforms traditional featurization techniques but also effectively bridges different
element types, providing a scalable solution for XANES
analysis across a wide range of materials. By first training a universal model at a fidelity with low computational cost, we demonstrate that ML models at computationally expensive fidelities can be efficiently developed by fine-tuning the universal model. The OmniXAS framework lays the groundwork
for future applications in high-throughput and real-time XAS analysis,
significantly reducing the reliance on computationally expensive simulations.

\section{Method}

\begin{figure*}[!thb] \centering
  \includegraphics[width=2\columnwidth]{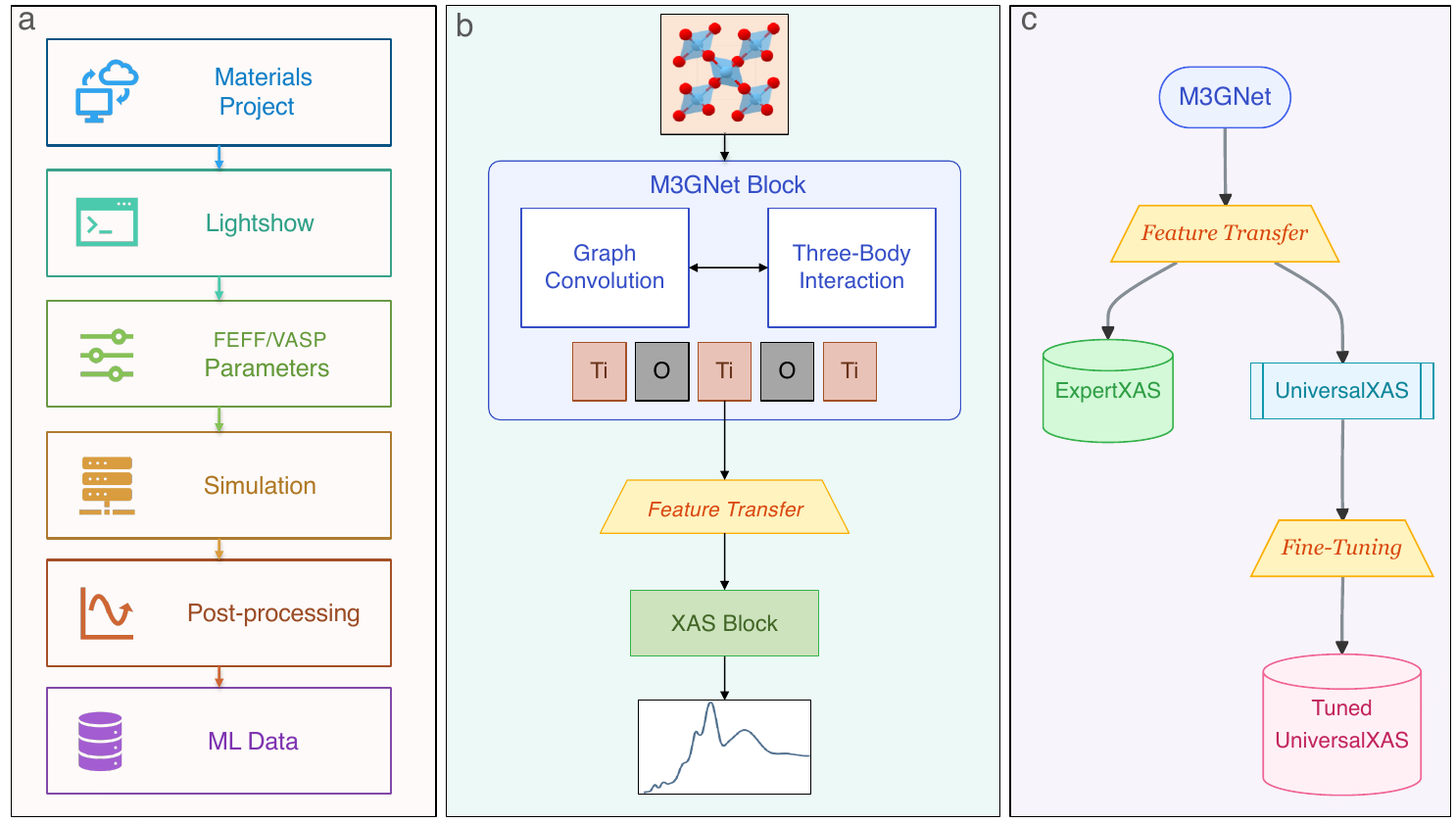}
  \caption{ Schematic of the OmniXAS framework.
    a) Data Curation:
    Structural data are sourced from the Materials Project, from which FEFF and VASP
    spectral input files are generated locally using Lightshow. FEFF and VASP spectral simulations are performed using these input files, and the results are screened and processed into machine learning-ready data.
    b) XAS Model: Materials structure information is processed through
    frozen M3GNet blocks using a series of graph convolutions with three-body
    interaction updates. From these operations, only the latent state at the
    node of the absorption site is passed into
    trainable neural networks for predicting site-specific XAS spectra.
    c) Cascaded Transfer Learning: The workflow diverges into
    two paths for training the XAS-block, resulting in three variants of
    XAS models. In one path, individual models for each data subset
    (ExpertXAS) are trained. In the other, a single model that predicts FEFF
    spectra for all elements (UniversalXAS) is developed. Further along, knowledge transfer
    from the UniversalXAS model is applied through fine-tuning, producing another
    set of specialized models for each subset that we call Tuned-UniversalXAS.
  }
  \label{fig:overview}
\end{figure*}

In this work, we demonstrate how a variety of deep learning models can be used
to accurately predict the XANES spectra of a large class of materials. Fig.~\ref{fig:overview} highlights our workflow of the OmniXAS framework, which includes
data acquisition, data curation, ML model architecture, and model training and
evaluation.

To create a database of paired structures and the corresponding XANES spectra, as shown in Fig.~\ref{fig:overview}a we pulled
structural data from the Materials Project~\cite{rankine2020deep}, generated spectral simulation input files  using the Lightshow package~\cite{carbone2023lightshow, Meng2024}, and performed spectral simulations, post-processing and data curation (Sec.~\ref{subsection:data acquisition and curation}). Two levels of theory were used in the spectral simulation: the real-space multiple scattering method implemented in FEFF9~\cite{rehr2010parameter} and the excited-electron core-hole potential method implemented in the Vienna Ab initio Simulation Package (VASP)~\cite{karsai2018effects}. This module ultimately produces the structure-spectrum pairs which are used as the training, validation and testing data in our ML models.

Next, multiple transfer learning ML models are developed
for site-spectrum prediction of each element, where the featurization of
the local structure of the absorbing site was taken from the
M3GNet~\cite{chen2022universal} via a feature transfer process
(Sec.~\ref{subsection:xas_hypothesis}-~\ref{subsection:transfer learning} and Fig.~\ref{fig:overview}b). 
We underscore that a universal
model can be developed using the data of all eight elements and subsequently fine-tuned on
specific element to further improve the model performance (Sec.~\ref{subsection:xas_hypothesis} and Fig.~\ref{fig:overview}c). Details of the
model architectures are explained in Sec.~\ref{subsection:featurization}.
Training and evaluation of these models, such as partitioning the data into
training, validation and testing splits, are described in
Sec.~\ref{subsection:split_train}.

\subsection{Data Acquisition and Curation} \label{subsection:data
  acquisition and curation}

Before ML models can be trained, structure and spectral data need to be sourced, curated, and standardized. We used the Lightshow Python
package~\cite{carbone2023lightshow,Meng2024} to pull all available Materials
Project structural data of pure metals and primary, secondary, and ternary oxides via Pymatgen~\cite{ong2013python} for Ti, V, Cr, Mn,
Fe, Co, Ni and Cu. FEFF9 and VASP spectral input files were also
written using Lightshow.

\subsubsection{Input File Generation}

FEFF calculations were performed in the real space, and the random phase approximation was used for core-hole screening. Self-consistent field and full multiple scattering lengths were set to 7 and 9~\AA, respectively.
In VASP spectral simulations, both ground state and core-hole excited state were calculated under the non-spin polarized configuration.
We used the GW-type pseudopotential, in order to obtain a good description of high-energy scattering states~\cite{Meng2024}. A full core hole was included in spectral simulations and the core electron was placed at the bottom of the conduction band. The \emph{k}-point mesh used in the Brillouin zone sampling is an important convergence parameter of the VASP spectral
calculation, which was determined using the effective crystal size
method~\cite{Meng2024} with the \emph{cutoff} parameter set at {24~\AA} in
Lightshow. The total number of bands ($n_b$) included in the spectral
calculation depends on system size and the chosen energy range. Here we
estimated $n_b$ based on the uniform electron gas model~\cite{Meng2024} with
the \code{e\_range} parameter set to 30~eV (20~eV) for Ti (Cu) in Lightshow.
The computational cost grows linearly with $n_b$. To make the high-throughput
calculations tractable, we only considered materials with $n_b$ less than 2200 in VASP
simulations.

\subsubsection{Removal of Unphysical Spectra and Outliers}

In the first data standardization step, site-specific spectra that failed
sanity checks were discarded, when the FEFF calculations
did not converge with the default input. Next we adopt a statistical approach to remove outliers.
Define $\vecmu^{(i)} = \left[\mu_1^{(i)}, \mu_2^{(i)}, ..., \mu_M^{(i)}\right]$ the $i$-th simulated spectrum (i.e., ground truth of the ML models) discretzied on $M$ energy grid points. The absorption coefficient at grid point $j$ averaged over $N_{tot}$ spectra is given by 
\begin{equation}
    \bar{\mu}_j = \frac{1}{N_{tot}}\sum_{i=1}^{N_{tot}} \mu_j^{(i)},
\end{equation}
and the corresponding standard deviation (without bias correction, i.e., using $N_{tot}$ instead of $N_{tot}-1$ in the normalization~\cite{van2011numpy,numpy121}) is given by
\begin{equation}
    \sigma_j = \sqrt{\frac{1}{N_{tot}} \sum_{i=1}^{N_{tot}} \left(\bar{\mu}_j - \mu_j^{(i)}\right)^2}.
\end{equation}
Any spectrum with $|\mu_j^{(i)}-\bar\mu_j|> c\,\sigma_j$ for any $j=1, ..., M$ is removed from the
database. In other words, spectra where the absorption coefficient lies outside of the
envelope set by a chosen multiple of the standard deviation are removed. For FEFF
and VASP spectra, we used standard deviation thresholds of $c=2.5$ and $5$,
respectively. These threshold choices were determined empirically
to remove the significant outliers observed in spectra heat maps
while retaining most of the data.

\subsubsection{Rescaling and Edge Alignment} \label{subsubsection:rescaling_and_alignment}

The raw output of the FEFF spectra under the default normalization was re-scaled to obtain the absolute cross section
with default broadening. The output of the VASP spectra is the imaginary part
of the macroscopic dielectric constant. To compare results between FEFF and
VASP, we converted the imaginary part of the dielectric constant to cross
section~\cite{Meng2024}. Raw VASP spectra were calculated purposely with a small
broadening of 0.05 eV. This allows one to further broaden the spectra in
post-processing as needed. We applied a 0.89 (3.46) eV core-hole lifetime
broadening to Ti (Cu) VASP spectra, such that the spectral resolutions
in FEFF and VASP are comparable.

Relative edge alignment was applied to VASP site-specific spectra using the
$\Delta$SCF method~\cite{england2011hydration,Meng2024}. After that, a constant empirical shift of 5108.59 (9492.80) eV
was applied to VASP Ti (Cu) spectra to align with reference experimental
spectrum of anatase TiO$_2$~\cite{carta2015x} (cuprite Cu$_2$O~\cite{gaur2009copper}).
In Sec.~\ref{subsection:transfer learning}, we need to further align
FEFF and VASP spectra, and an additional constant
shift of $-5.5$ ($7.0$)~eV was applied to VASP Ti (Cu) spectra such that the average spectra in
VASP align with its FEFF counterparts. After post-processing, we consider the same
energy range of 35 eV for all the spectra spanned on a uniform energy grid of
$\delta E=0.25$~eV from different starting energies determined empirically for each element (see
Supplementary Table S1). This
results in a 141-dimension vector that represents each spectrum. Values of the relative energy $\Delta E$ are defined on this uniform energy grid as $\Delta E_j = \delta E \times j$, 
where $j$ is the grid index.

\begin{figure}[htbp] \includegraphics[width=3 in]{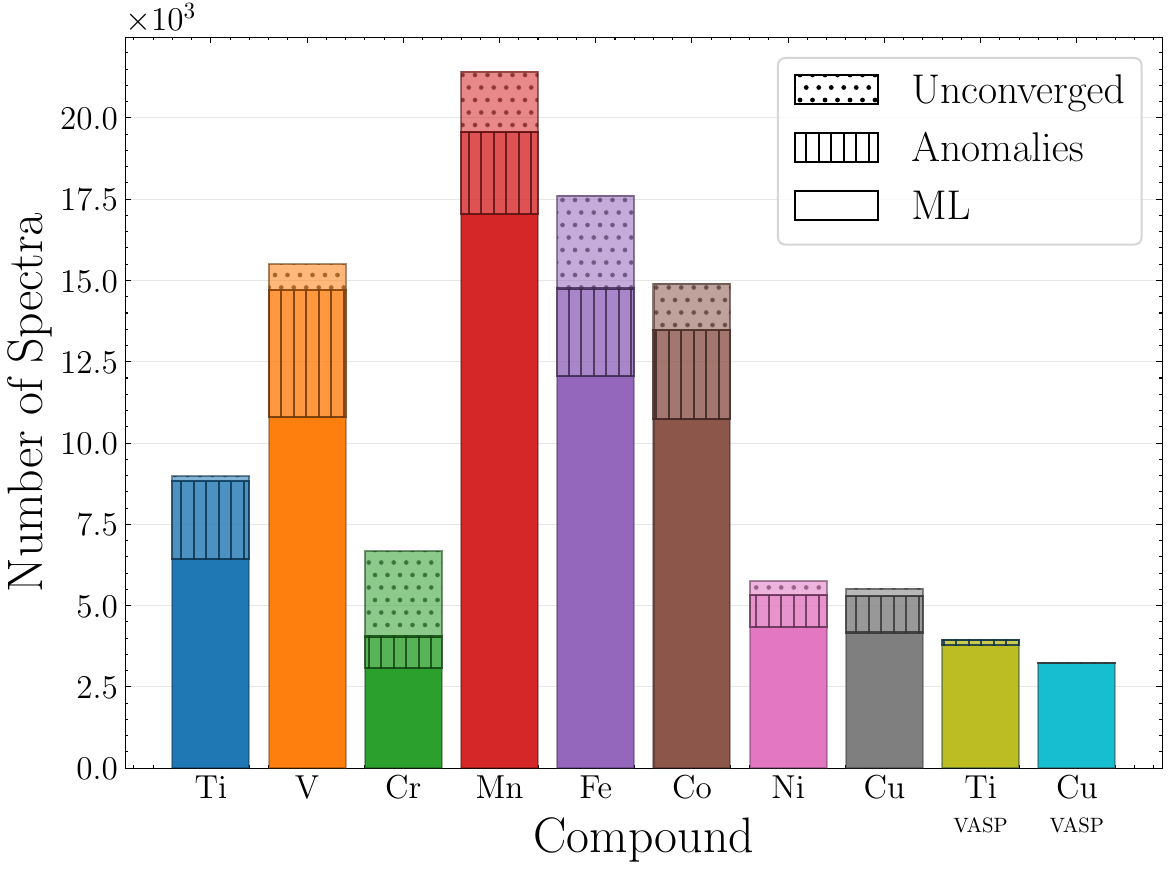}
  \caption{
    Number of spectra generated for each element from FEFF and VASP
    simulations. The highest bars represent the total number of spectra, with
    partitions indicating the portions removed during different data cleaning stages
    (unconverged and anomalies) and the remaining ML-ready
    data. \label{fig:datasizes}}
\end{figure}

The final dataset after curation and standardization consists of 75,691 site-specific spectra
across all eight elements from two simulation methods, which we refer to as the 3d transition metal XANES dataset, ``3dtm\_xanes\_ml2024". The detailed data size breakdown is summarized in Fig.~\ref{fig:datasizes} and Supplementary Table S2.

\subsection{XAS Prediction Hypotheses} \label{subsection:xas_hypothesis}

We propose two key hypotheses for XAS prediction that guide the rest of the methodologies for ML model design:

\begin{itemize}
    \item \emph{Feature Transfer Hypothesis:} The structural features in ML models trained for
 materials inter-atomic potentials provide sufficient transferable knowledge for XAS prediction models. 
    \item \emph{Universal XAS Model Hypothesis:} A universal XAS model, trained on diverse elements, can extract core knowledge of the structure-spectrum relationship that is also transferable to improve element-specific predictive models.    
\end{itemize}

These two hypotheses lead us adopt a cascaded transfer learning approach when developing our ML models.

\subsection{Transfer Learning} \label{subsection:transfer learning}

Transfer learning draws inspiration from human cognition, where knowledge
acquired from one task can be applied to another ~\cite{torrey2010transfer}. This knowledge transfer
approach is adopted in machine learning by leveraging a pre-trained source
model to improve the training efficiency and performance of a distinct target
model.

Like in human learning, transfer learning is effective when the learning tasks
of source and target models are related, if not identical. This method has
found success in many fields~\cite{zhuang2020comprehensive}, such as computer vision~\cite{duan2012learning,kulis2011you,zhu2011heterogeneous}, natural language
processing~\cite{zhou2014hybrid,prettenhofer2010cross,zhou2014heterogeneous}, and speech recognition~\cite{xia2011pos}, offering benefits like reduced training
time, improved performance, and better generalization. Due to the high
computational cost to generate XAS spectra using first-principles band-structure
codes, constructing large XAS datasets with different levels of theory remains
a bottleneck. Effective transfer learning provides a practical route to leverage data from different fidelities and significantly enhance the performance of XAS ML models.

Transfer learning is categorized based on approaches used, level of
similarity of source and target tasks as well as that of input (domain)
data~\cite{panyang2010}. In this paper, we focus on two of such transfer learning
categories: \textit{inductive-transfer-learning} and \textit{domain-adaptaion},
for which we use the \textit{feature-transfer} and \textit{fine-tuning}
approach, respectively.

\subsubsection{Inductive Transfer Learning via Feature-Transfer} \label{section:inductiveTL}

Our methods to investigate the ``Feature Transfer Hypothesis"  fall under the
category of \textit{inductive transfer learning}~\cite{panyang2010} in the ML
terminology, as the tasks between source and target models, while distinct,
share similarities in predicting physical properties of materials based on
their graph representations. To harness this potential, we selected a
pre-trained interatomic potential model called M3GNet~\cite{chen2022universal} as our transfer source, and hypothesize that it contains
valuable, transferable knowledge for XAS prediction.

Among several approaches that can be taken for inductive transfer learning, we
implemented \textit{feature-based transfer learning}. This methodology
is guided by the principle that representation of input in early
layers of neural networks should be similar across models trained for different
(but similar) tasks. One important note to this approach is that XAS prediction
is focused mostly on local properties compared to M3GNet, which predicts both
global properties (total energy and stress) as well as local properties (atomic
forces). This intricacy suggests that only a subspace of the full latent
representation of M3GNet is needed in knowledge transfer onto XAS prediction. Our
design and implementation details for this model adaptation are described in
Sec.~\ref{subsection:featurization}.

\subsubsection{Domain Adaptation via Fine-tuning}\label{subsubsection:hypothesisUniversalXAS}

To further investigate the ``Universal XAS Model Hypothesis", we complement our inductive transfer learning approach with domain adaptation. This method is particularly effective when the task remains consistent but the input data distribution shifts—a common scenario when predicting XAS across diverse elements. Our domain adaptation strategy is implemented in two stages: a) the development of a universal XAS model, and b) fine-tuning universal models with element-specific XAS data.

We begin by developing a universal XAS prediction model
designed to operate across eight 3d-transition metals. This model is trained on
an extensive dataset generated from FEFF simulations, encompassing a wide array
of structural and site configurations. The fundamental premise underlying this
approach is rooted in the intuition that while XAS spectra exhibit
element-specific characteristics, the core physical principles governing X-ray
absorption remain sufficiently similar across elements, which lead to general spectral
trends beyond a single element.

For example, since 1920 it has been established that the edge position in the
XANES spectra strongly correlates with the oxidation state of the absorbing
atom~\cite{bergengren1920rontgenabsorption,kunzl1932linear,pauling1960nature,suchet1965physical}.
The absorbing atom can exhibit a positive (negative) edge
shift depending on whether it is oxidized (reduced), which can be understood as
the shielding effects of valence electrons on the core-electron energy levels.
Another important trend in 3d transition metal K-edge XANES is the relation
between the pre-edge features and the cation coordination
number and local symmetry~\cite{farges1997ti,yamamoto2008assignment}. This relation is governed by the
dipole-forbidden $1s \rightarrow 3d$ transition in the pre-edge of the 3d
transition metal. The pre-edge intensity vanishes in a perfect octahedron, but
becomes more evident, when the inversion symmetry is broken that allows the
mixing of $p$ and $d$ orbitals. Overall, the pre-edge feature is pronounced in
early 3d transition metals (e.g., Ti)~\cite{farges1997ti}, but becomes less
obvious in late 3d transition metals~\cite{carbone2019classification}, as
filled 3d orbitals in the latter block the $1s \rightarrow 3d$ transitions.
Liang~\emph{et al.} constructed a rank-constrained adversarial autoencoder to
disentangle individual structure-spectrum relationship in the same material
space as this study~\cite{liang2023decoding}. In addition to oxidation state and coordination
number, the authors identified distinct spectral features associated with a set
of less well-known local structure descriptors, such as oxygen coordination
number, standard deviation of nearest neighbor (NN) bond angles, and minimum
O-O distance on the edges of the NN polyhedron~\cite{liang2023decoding}. The
distinct spectral features are not limited to edge position and pre-edge
intensity but also include more subtle ones,  such as main edge position 
and intensity, post-edge position and intensity and curvature of the pit, based on the terminology of 
Guda~\emph{et al.}~\cite{guda2021understanding}.

Given the domain knowledge on spectral trends beyond single element, we believe
that ML models can learn to capture not just element-specific features, but also
essential and transferable knowledge for XANES prediction, creating a robust
foundation for element-specific adaptations.
Following the training of the universal XAS model, we perform
targeted fine-tuning for specific elements of interest. This process allows the
model to adapt its learned representations to the unique spectral signatures of
individual elements while preserving the broader understanding gained from the
diverse training set.

Based on the description above, we introduce the following model names used
throughout the rest of the text as shown in Fig.~\ref{fig:overview} as well. We
also collectively refer them as \textit{XASModels}.

\begin{itemize}
    \item \textit{ExpertXAS}--- These models are trained to predict site-specific spectra
        for a particular element using only the corresponding subset of data from
        either FEFF or VASP simulations.
    \item \textit{UniversalXAS}--- This model is trained to predict spectra for all eight
        3d-transition metals using the complete set of data from FEFF simulations.
    \item \textit{Tuned-UniversalXAS}--- These models are created by fine-tuning the
        UniversalXAS model using only the corresponding subset of data from either
        FEFF or VASP simulations.
\end{itemize}

\subsection{XASModels} \label{subsection:featurization}

In this section, we introduce our model for site-specific XAS prediction.
As outlined in Sec.~\ref{section:inductiveTL}, we employ a feature-based transfer learning
approach, leveraging the pretrained M3GNet model. We first briefly summarize the core
components of M3GNet. Then, we explain how XASModels are constructed by combining
two key elements: frozen feature extraction layers from M3GNet and new
trainable neural network blocks designed specifically for XAS prediction, which
we call \textit{XAS blocks}.

\subsubsection{M3GNet block} \label{subsubsection:m3gnetBlock}

M3GNet is a graph deep learning model designed to predict the inter-atomic
potential, forces, and stresses from the material graph
representation~\cite{chen2022universal}. It was trained on a comprehensive
dataset from the Materials Project, comprising over 17 million entries and
encompassing 89 elements from the periodic table. 
The efficiency of M3GNet was demonstrated by the study that predicted almost 2 million
stable materials from a total of approximately $32$ million candidate
structures~\cite{chen2022universal}.

Inspired by the significance of three-body interactions in inter-atomic potentials, M3GNet distinguishes itself in its design from other graph deep learning models by embedding three-body interaction within the graph edge features. Specialized ``many-body to bond'' neural blocks are used to update latent features, incorporating three-body interactions among atoms based on their precise atomic coordinates. These ``many-body to bond'' blocks are sandwiched between graph convolution layers. A pair of these ``many-body to bond'' and graph convolution layers is collectively referred to as an \emph{M3GNet block}. 

As information flows through M3GNet, it passes through three of these M3GNet blocks before final decoding using the readout block. We use the latent space representation from the \textit{final} M3GNet block to encode structural information for XAS prediction.

\subsubsection{Transfer-Features}   \label{subsubsection:tranferFeatures}

The M3GNet architecture is designed to predict both global and local material properties,
with its latent space at each layer containing information relevant to both.
In contrast, XASmodels focus solely on local properties. This
fundamental difference needs to be taken into account when attempting to leverage
M3GNet's capabilities for XAS prediction through transfer learning.
The key lies in identifying an appropriate subset of M3GNet's latent space that
strikes a delicate balance. This subset must be localized, capturing the
specific site of interest for XAS prediction to ensure specificity of results.
Simultaneously, it needs to be comprehensive enough not to discard potentially
valuable information (e.g., intermediate range structural information beyond the first near-neighbor shell) that is known to contribute to XAS features. 
A significant complexity in this process stems from M3GNet's ``black box" nature – a common issue in machine learning where advanced models have internal mechanisms that are not easily interpretable ~\cite{krause2016interacting,castelvecchi2016can}. This lack of transparency makes it difficult to identify which aspects of the model's latent space are most crucial for accurate XAS prediction. To address these issues, we adopt the following approach for latent space selection: we extract node features from the graph after final graph convolution layer and just before M3GNet's readout block, focusing on the site where the absorption element information was initially encoded. 
Our rationale is that M3GNet's training on both local and global property prediction may have naturally produced a latent space that can be segmented into clusters containing rich local information surrounding the sites where atomic data was initially encoded, while still preserving some global context relevant to XAS prediction.
While intuitively appealing, it is important to note that the
effectiveness of this approach needs to be validated, a task we undertake in
Secs.~\ref{subsection:analysisfeatureTransfer} and~\ref{subsection:resultTransferFeature}.

The M3GNet block can be represented as a function that takes
in the material structure's graph representation $\mathcal{S}$ 
and generates features in latent space, i.e.
\begin{equation}
    \mathbf{H}_0 = f(\mathcal{S}) \in \mathbb{R}^{n \times 64}.
\end{equation}
Here, $f: \mathcal{S} \rightarrow \mathbb{R}^{n \times 64}$ represents the M3GNet function that transforms an input structure $\mathcal{S}$ containing $n$ atoms into a matrix $\mathbf{H}_0$. Each row of $\mathbf{H}_0$ encapsulates the features around the location where an individual atom $i$ was initially encoded in the graph.

Our focus is the site-specific spectrum, so we extract the feature vector for the $i$-th absorbing atom as:
\begin{equation}
    \mathbf{h}_0^{(i)} \in \mathbb{R}^{64}.
\end{equation}
This vector, corresponding to the $i$-th row of $\mathbf{H}_0$, serves as the input to the subsequent MLP layers and is henceforth referred to as \emph{transfer-features}.

\subsubsection{XAS block} \label{subsubsection:xasBlock}

The XAS block module harnesses the power of MLPs to transform M3GNet-extracted transfer-features into accurate XAS spectra. At their core, MLPs are composed of interconnected neuron layers that sequentially process and refine input data. As transfer-features flow through the XAS block, each layer applies complex nonlinear transformations, enabling the network to capture intricate relationships between transfer-features and resulting XAS spectra. The universal approximation theorem of MLPs, a fundamental principle in machine learning, mathematically establishes that a properly constructed XAS block can arbitrarily closely approximate any continuous function mapping transfer-features to XAS spectra, provided sufficient network depth and optimal parameter configuration.

Each layer of the MLP applies a carefully chosen sequence of operations: linear transformation, batch normalization, activation, and dropout. This sequence maintains the network's generalization capabilities and enhances learning. For hidden layers ($l = 1, 2, ..., L-1$), the computation is defined as:

\begin{multline}
    \mathbf{h}_l = \mathrm{Dropout}_l(
            \mathrm{SiLU}(
                \mathrm{BatchNorm}_l(
                    \mathrm{Linear}_l(\mathbf{h}_{l-1})
                )
            )
        ) \\ \forall \quad l \in \{1, 2, ..., L-1\}.
\end{multline}
This formulation encapsulates the layer-wise transformation of features, with each operation serving a specific purpose in the learning process.
The final output layer differs slightly to others by using Softplus, 
\begin{equation}
    \hat{\bm{\mu}} = \mathbf{h}_L = \mathrm{Softplus}(\mathrm{Linear}_L(\mathbf{h}_{L-1})).
\end{equation}
Here, $\hat{\bm{\mu}} \in \mathbb{R}^{141}$ represents the predicted XAS spectrum across 141 energy grid points.
Each component of the MLP serves a specific purpose in the learning process, contributing to the model's overall effectiveness. Further details of the MLP can be found in the Supplementary Materials.

The flexibility of the MLP allows for extensive optimization of both architectural and training parameters. Key architectural variables include network depth (number of layers) and width (neurons per layer), which directly impact the model's capacity to learn complex patterns. Crucial training hyperparameters, including batch size, influence optimization stochasticity and generalization ability.

To efficiently navigate this vast hyperparameter space, we employ Bayesian optimization technique implemented in Optuna Package~\cite{optuna}. This approach systematically explores potential configurations, using validation set performance to guide the search for optimal combinations of architectural and training parameters. The exploration of the search space and the resulting optimized configurations for all XASModel instances are detailed in Supplementary Table S3-S4.

\subsection{XASModel Training} \label{subsection:split_train}

In this section, we describe our methodology for training XASModels. We introduce
our data partitioning scheme, which gives more consistent split sizes
across data sets of different sizes and material distributions. This also
ensures that there is no material information leakage between them. We then
detail the training steps taken to ensure robust model performance.

\subsubsection{Balanced Material Splitting}

For training, we first divide the data into training, validation, and test
sets, used respectively for model fitting, hyperparameter tuning, and final
performance evaluation. The goal of our XAS prediction models is to be
generalizable to unseen \textit{materials}, not just unseen \textit{absorbing
  sites}. To achieve this, we employ materials splitting, where all XAS spectra
from a single material's absorption sites are grouped together, ensuring they 
always fall into the same split. This
ensures no data from any single material appears across multiple splits (e.g.,
a material with two unique absorbing sites, in which one site's data is in the training set and 
the other site in the testing set), preventing data leakage and
overly optimistic performance estimates. We call this approach \textit{materials splitting}, which provides a more stringent assessment of a model's 
generalizability to new compounds.

Materials splitting raises an issue in maintaining consistent proportions
across training, validation, and testing sets. While we target specific ratios
(e.g., 80/10/10), the uneven distribution of XAS data across materials
complicates achieving these exact proportions. Some materials in our dataset
have more unique site-XAS spectra than others, creating
an inherent imbalance. Consequently, when splitting by material, the actual
data proportions in each set may slightly deviate from the intended ratios. This
discrepancy is especially noticeable in datasets with a wide variance in the number of site-XAS
spectra per material or in smaller datasets. Such imbalances can potentially
impact the consistency of model training and evaluation, necessitating careful
consideration during the data preparation phase.

We designed a heuristic algorithm to address the issue of attaining desired
proportions in the splits while maintaining materials splitting. The
algorithm takes a greedy approach when assigning unique materials to groups. It
begins by sorting the unique materials by their number of unique sites in the
descending order, then iteratively assigns them to groups. This sorting scheme ensures
that materials with the largest number of sites are handled first, allowing for
more effective balancing of the splits later on. For each unique material, the
algorithm calculates how adding it would affect each group's proximity to its
target size. It then assigns the material to the group where this addition
would most effectively reduce the gap between current and target sizes. This
process continues until all the materials are allocated.

\subsubsection{Training}

We trained the XASModels with the Adam optimizer~\cite{kingma2014adam} using the mean squared error (MSE) loss function, 
\begin{equation}  
J \equiv \frac{1}{N} \sum_{i=1}^N J^{(i)},
\end{equation}
where
\begin{equation} \label{eq:mse}
\quad J^{(i)} = \frac{1}{M} \sum_{j=1}^M (\mu_j^{(i)} - \hat{\mu}_j^{(i)})^2
\end{equation}
and $N$ is the number of spectra in the training set. $\mu_j^{(i)}$ and $\hat{\mu}_j^{(i)}$ represent the ground truth and the predicted absorption coefficient, respectively, at the $j$-th energy grid point of the $i$-th target spectrum.

Before the start of each training run, we used an automated search~\cite{learningratefinder} to determine a good initial learning rate. To prevent
overfitting of models, we terminated training whenever the minimum
validation loss did not improve for 50 consecutive epochs. The maximum number
of epochs was set to 1000, which was verified to be sufficient to allow for
convergence in training of all model variants across all element datasets.

When generating Tuned-UniversalXAS models, we face two distinct fine-tuning scenarios. The first involves fine-tuning the UniversalXAS model with element-specific FEFF datasets, where the training data is a subset of what was used to train the original UniversalXAS model. The second involves fine-tuning with VASP datasets, which introduces both a different fidelity and variance, demanding more substantial model adaptation. Given these different adaptation requirements, we systematically evaluated dropout regularization using two configurations: maintaining the original dropout rate of 0.5 (consistent with ExpertXAS and UniversalXAS models) and disabling the regularization by setting it to 0. This approach allows us to determine the optimal approach for each type of Tuned-UniversalXAS model.

\section{Preliminary Analysis and Data Exploration}

Data exploration serves as an important indicator of the overall data quality
of the dataset and helps to uncover potential connections between features and
targets. This is a necessary step before training ML models.

The ML-data features and targets frequently exist in high-dimensional spaces,
rendering data exploration challenging. Our work, for instance, employs
64-dimensional feature-transfer vectors and 141-dimensional target vectors for
each data point. 
The number of spectra in dataset for each element range from 3,067 (Cr) to 17,052 (Mn).
To address the challenge of high data
dimension, we use two approaches for preliminary data analysis and exploration:
a) direct visualization of the spectral data using heatmaps and b)
dimensionality reduction-assisted analysis of both feature and target data.

We employ Uniform Manifold Approximation and Projection (UMAP) to visualize our
feature and target data~\cite{mcinnes2018umap}.
UMAP is a powerful dimensionality reduction technique that effectively preserves the topology of both local and global data structures. It first constructs a high-dimensional representation using a fuzzy topological framework, specifically a weighted $k$-neighbor graph where edge weights represent the probabilities of connections between points. Then a low-dimensional representation is constructed with the same method, which is optimized by minimizing the cross-entropy between the high- and low-dimensional fuzzy representations through a stochastic gradient method. Due to its stochastic nature, UMAP projections are inherently non-deterministic and can vary based on parameters like the number of neighbors and minimum distance. While these variations can lead to different details in clustering outcomes, even a single projection that reveals clear clustering provides strong evidence for the existence of clusters in the original high-dimensional space. UMAP often uncovers intricate patterns and cluster structures that may be obscured in the original data~\cite{mcinnes2018umap}.

\subsection{Spectra Visual Inspection} \label{subsection:spectra}

\begin{figure*}[!thb]
  \centering
  \includegraphics[width=\linewidth]{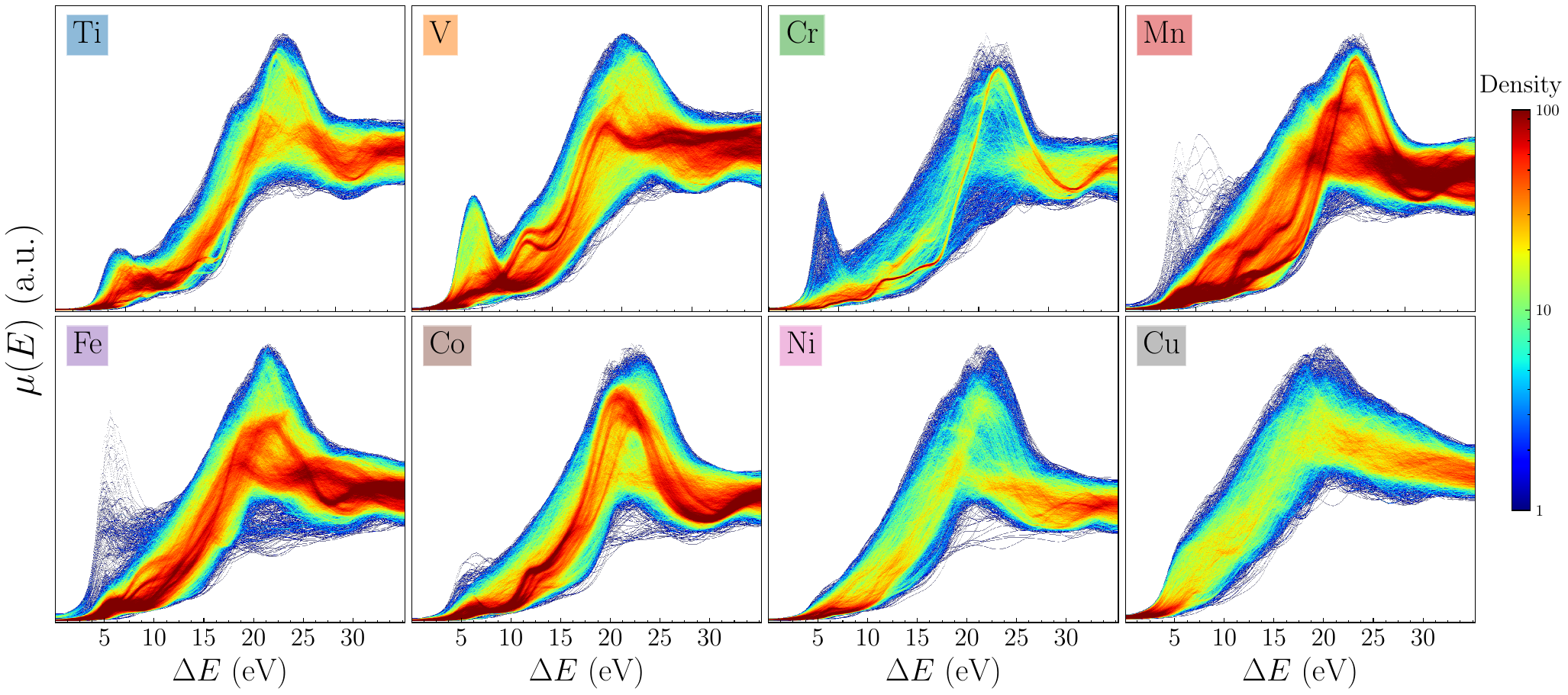}
  \caption{
    Heatmaps of FEFF XANES spectra. 
    Color represents the density of spectral features. \label{fig:FEFF-ML-data}
  }
\end{figure*}

\begin{figure}[!thb]
  \centering
  \includegraphics[width=\linewidth]{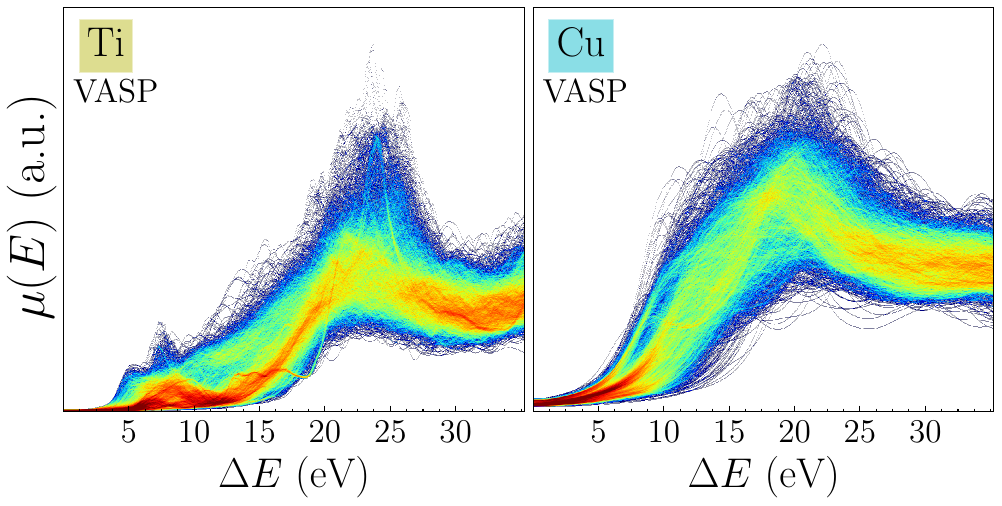}
  \caption{
    Heatmaps of VASP XANES spectra. Color scheme is the same
    as in Fig.~\ref{fig:FEFF-ML-data}. \label{fig:VASP-ML-data}
  }
\end{figure}

We proceed by direct visual inspection of the spectra database with focus
first on the high-density region (dark red) of the entire FEFF database in
Fig.~\ref{fig:FEFF-ML-data}. Ti and V XANES exhibit clearly
visible pre-edge peaks. The pre-edge peak intensity decays when moving towards
late 3d transition metals, consistent with previous
observations~\cite{carbone2019classification}. Mn and Fe XANES show a much
wider range in edge location than others, consistent with multiple commonly observed
oxidation states. Mn XANES shows at least three edge positions as families of dark red curves and Fe XANES
shows at least two with smaller spacing than Mn. Correspondingly, Mn and Fe
spectra exhibit a broader spectral variation or data complexity than others. In
contrast, Ni and Cu XANES show a relatively simple spectral shape with one
dominant peak. Regarding Ti and Cu VASP database in Fig.~\ref{fig:VASP-ML-data},
overall the spectral shape in
VASP are similar to that in FEFF. However, the low density region (dark blue)
in VASP has larger variations than FEFF in both elements.

\subsection{Transfer-Features Clustering} \label{subsection:analysisfeatureTransfer}

\begin{figure*}
  \includegraphics[width=\linewidth]{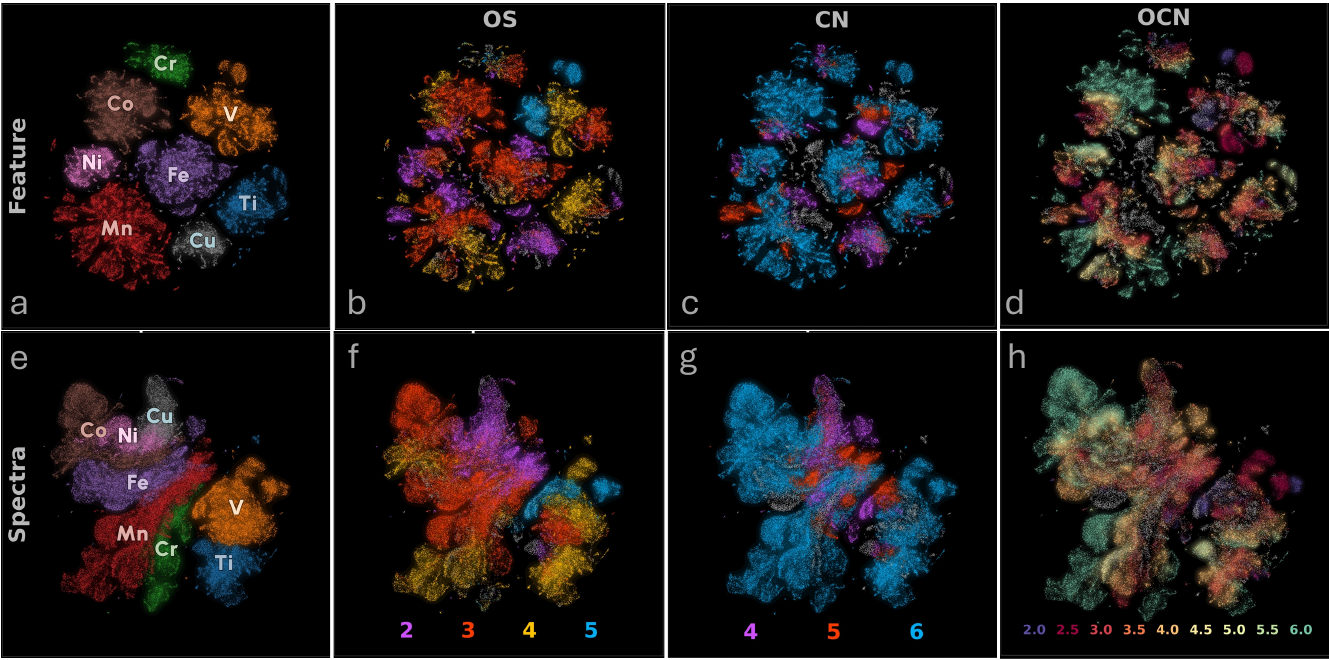}
  \caption{
    UMAP plots of M3GNet featurization (a-d) and spectra (e-h)
    colored by absorbing element type (a and e), oxidation-state
    (OS) (b and f), coordination-number (CN) (c and g), and oxygen-coordination-number (OCN) (d and h). The
    oxygen-coordination-number is rounded to the nearest half integer for visualization
    purposes. The gray points represent those in which the physical descriptor
    can not be determined.
  }
  \label{fig:umapFEFF}
\end{figure*}

We select the node features of the absorbing site within the M3GNet latent space for transfer-feature extraction, as detailed in Sec.~\ref{subsubsection:tranferFeatures}. While this choice is intuitively appealing, it necessitates careful examination due to the \textit{feature smoothing} phenomenon. Feature smoothing occurs when node features gradually become similar with each other in deeper layers of the model~\cite{cai2020gnnsmoothing,chen2020measuring}.
This happens because the deeper convolutional layers increasingly expand the
effective receptive field from which the features are being aggregated. In the
case of M3GNet, feature-smoothing could potentially dilute local structural information when it passes through the M3GNet block, thereby hindering the site-specificity of the predicted XAS and model performance.

We provide empirical evidence that feature dilution in the
M3GNet block is sufficiently mild. We demonstrate this by visualizing the
transfer-feature in low dimensions with UMAP and identifying 
clusters that correspond to local physical
descriptors. Specifically we use the oxidation state (OS) of the absorbing cation,
the number of nearest neighbors of the absorbing cation atom, i.e. the
coordination number (CN), and the average number of nearest neighbors of oxygen
atoms in the first shell of the absorbing cation, i.e. the oxygen
coordination number (OCN)\footnote{For example, consider an absorbing Ti atom
with 2 nearest neighbor Zn and two nearest neighbor O atoms. If one of the O
atoms has 2 nearest neighbors, and the other O atom has 4 nearest neighbors,
then the OCN value for the absorbing Ti atom is 3.} as the labels.

We plot two-dimensional UMAP of all transfer features in the FEFF ML dataset in
Fig.~\ref{fig:umapFEFF}a-d, colored by element type and local physical
descriptors (OS, CN, and OCN). In doing so, we make two
key observations as summarized below that provide compelling evidence that the
latent-subspace-based transfer features can be used for XAS prediction without
concern for excessive feature smoothing inside the M3GNet block.

\begin{itemize}
    \item \emph{Element Clustering}--- 
        The transfer-feature UMAP in Fig.~\ref{fig:umapFEFF}a
        shows distinct clusters based on the absorbing element type.
        The figure clearly demonstrates that these clusters map very well to the
        absorbing element labels from which they originally assigned (marked by
        different colors). This provides strong evidence that transfer features are
        sufficiently distinct among different absorbing elements, despite some moderate
        feature smoothing.
    \item \emph{Local Descriptor Clustering}--- The UMAP plots, when colored
        by local physical descriptors as shown in Fig.~\ref{fig:umapFEFF}b-d, reveal
        consistent sub-groupings within element clusters. This observation is
        consistent across different physical descriptors, indicating that transfer
        features are sufficiently rich to capture local structure information, both
        at the level of the absorber element type and the sub-level of physical
        descriptors (OS, CN and OCN) within each element.
\end{itemize}

\subsection{Spectra Clustering} \label{subsection:spectra clustering}

Now we turn to the UMAP clustering of the FEFF spectra as shown in Fig.~\ref{fig:umapFEFF}e-h. In prior studies~\cite{carbone2019classification,carbone2020machine,ghose2023uncertainty}, spectral analysis through dimensionality reduction, such as principal component analysis (PCA), is typically performed on datasets of a single absorbing element. This is because XANES is an element-specific technique, and one can trivially distinguish the K-edge XANES of 3d transition metals from their absolute edge position energy. In this study, we apply UMAP to the entire FEFF database (Ti -- Cu) defined on a relative energy scale, thereby removing the information regarding absolute edge positions. This approach allows the UMAP analysis to focus exclusively on spectral shapes. By visualizing the \textit{entire} FEFF dataset in a single UMAP plot, we can identify relationships and trends that transcend individual elements, highlighting how spectral features correlate with broader material properties.

As shown in Fig.~\ref{fig:umapFEFF}e, distinct clustering patterns form for
each element. In addition, elements next to each other in the periodic table
stay close together: early 3d transition metals (Ti, V, Cr and Mn) and late 3d
transition metals (Fe, Co, Ni and Cu). However, there is not a clear separation
between early and late 3d transition metals. The spectral similarity between
different elements is supported by the observation in the raw spectra in
Fig.~\ref{fig:FEFF-ML-data}, e.g., the pronounced pre-edge in Ti, V and Cr, and
the single peak feature in Ni and Cu. The distinction between different elements observed in the spectral UMAP may arise from several factors, including differences in core-hole lifetimes and absorber atom phase shifts, as well as variations in scattering path lengths caused by differing bond lengths under the same local symmetry.

Further spectra clustering analysis was performed by labeling the UMAP pattern
with physical descriptors (OS, CN and OCN) as shown in Fig.~\ref{fig:umapFEFF}f-h. Within each element, one can see
sub-level clustering with respect to the physical descriptors. For example,
there is a dominant Ti$^{4+}$ cluster (yellow) accompanied by a much smaller Ti$^{2+}$
cluster (purple), while the Ti$^{3+}$ cluster (red) is less significant and more scattered.
This is consistent with the fact that the Ti database is dominated by Ti$^{4+}$
cations. Furthermore, there is a clear distinction in Mn and Fe 
clusters with respect to OS (2+, 3+ and 4+), reflecting the clear trend in edge positions in the raw spectra in
Fig.~\ref{fig:FEFF-ML-data}, because edge position is a well established proxy of ion OS. Similarly, sub-regions of CN can be seen within each cluster, e.g. 4-, 5- and 6-coordinated  Ti and V with 6-coordinated motifs being most prevalent,
consistent with previous work based on PCA
analysis~\cite{carbone2019classification}.
OCN is a more subtle physical
descriptor, which is correlated to main peak
intensity~\cite{liang2023decoding}. However, such trend is not obvious in the
raw spectra, as it is entangled with spectral variations caused by other
physical descriptors~\cite{liang2023decoding}. It is quite remarkable that UMAP
can in fact form distinct clusters of OCN, although its pattern is more complex
than OS and CN.

As shown by the above analysis, 3d transition metal K-edge XANES encodes key
information of the local chemical environment (e.g., OS, CN and OCN). To train
an ExpertXAS model, the structure features are required to capture the
variation of these physical descriptors. The success in M3GNet feature transfer
in identifying such variations in Fig.~\ref{fig:umapFEFF} is a necessary
condition of the validity of the feature transfer hypothesis,
establishing the effectiveness of M3GNet feature transfer for predicting
XANES spectra.

\subsection{Evidence of Universality} \label{subsection:analysisHintsOfUniversality}

The UMAP analysis in Fig.~\ref{fig:umapFEFF} also provides compelling evidence
that XANES prediction can take advantage of the spectral features common to all absorbing 
elements (Ti -- Cu), setting the stage for the UniversalXAS model.

When we color the UMAP pattern of the entire FEFF dataset based on the
absorbing element type, distinct clusters are formed corresponding to each
element, in both the featurized structure space and spectral space. 
Nevertheless, closer inspection reveals
that certain subset of spectra of one element can be more similar to spectra in
different element clusters than to those within the same element. For example,
spectra at the top right of the elongated Mn cluster in Fig.~\ref{fig:umapFEFF}e are closer
to the neighboring spectra in the Fe and Cr clusters than those at opposite
ends of the Mn cluster. Similar observations can be made for other element
clusters as well. From domain knowledge, we know
that fingerprints of OS (e.g., edge position) and CN (e.g., pre-edge peak
position and intensity) could serve as good candidates of the across-element
trends. In addition, the analysis of the rank constrained adversarial
autoencoder~\cite{liang2023decoding} also made the strong case for
across-element spectral trends associated with additional physical descriptors,
e.g., OCN, following a spectral disentanglement procedure.

To further investigate the cross-element spectral trends, we revisit the
spectral UMAP analysis labeled by local physical descriptors (OS, CCN, and OCN)
in Fig.~\ref{fig:umapFEFF}f-h. We can see distinct sub-regions of physical
descriptors (distinguishable by different colors) within each element cluster
often coalesce into larger clusters across the boundary of elements. This
coalescence is particularly pronounced in the OS UMAPs  (e.g., OS=2: Cu$^{2+}$,
Ni$^{2+}$ and Fe$^{2+}$ in pink color; OS=4: Ti$^{4+}$ and V$^{4+}$ in yellow color), underscoring the strong
similarity in the spectral feature of different 3d transition metals at the
same OS. Similar observations can also be made for CN and OCN, except that the
across-element coalescence is not as concentrated as OS. These qualitative
visualizations indicate that local chemical environments, in addition to the
absorbing element type, are among the primary characteristics that influence
XANES prediction and their trends can cross element boundaries.

\section{Results} \label{section:results}

\subsection{Performance Metrics} \label{subsection:resultPerformanceMetric}

When evaluating XASModels, it is imperative to fairly evaluate and compare the
performance of models trained on different subsets of
the spectra dataset of varying scales associated with the level of theory (the
multiple scattering method in FEFF and the core-hole pseudopotential method in
VASP) and element type (absolute cross-section of eight 3d transition metals).
To enable this comparison, we introduce a relative performance metric, $\eta$,
with intention to provide a single, interpretable metric that is robust to
outliers and allows for meaningful comparisons.

Our primary interest lies in spectral prediction accuracy, which we quantify
using the MSE [c.f. Eq.~\ref{eq:mse}] for each individual spectrum predicted by a
model. To represent a model's overall performance for a set of spectra, we
introduce the Median Spectral Error, denoted as $\xi$,
\begin{equation}
  \xi = \mathrm{Median}(\{J^{(i)}\}_{i=1}^N),
\end{equation}
where $J^{(i)}$ is the MSE of the $i$-th spectrum and
$N$ is the number of spectra in the test set. We choose the median of 
error, instead of other
representative statistics like the mean, because different models
have different error distributions which are often skewed and
contain outliers. The median provides robustness against these factors.

To address the issue of varying spectral scales in different spectra
sub-datasets, we normalize $\xi$ to bring it to a similar scale, yielding the
relative performance metric $\eta$,
\begin{equation}
  \eta = \frac{\xi_\mathrm{baseline}}{\xi}.
  \label{eq:eta}
\end{equation}
Here, $\xi_\mathrm{baseline}$ is derived from a simple baseline model that
always predicts the mean spectrum of the training set,
\begin{equation}
  [\mu_\mathrm{baseline}]_j = \frac{1}{N_{train}} \sum_{i=1}^{N_{train}} \mu_j^{(i)}.
\end{equation}
The $\eta$ metric provides an intuitive measure of model
performance, enables meaningful comparisons across sub-datasets with different
spectral scales, and preserves the robustness to outliers inherent in $\xi.$

By definition, $\eta_\mathrm{baseline} = 1$. The $\eta$ metric is unbounded
above 1, with higher values indicating better performance. 
While $\eta$ effectively captures the overall performance, it does not provide
a detailed description of the model performance. Therefore, we further
complement this metric with examination of error distributions and
visual inspection of predicted spectra, ensuring a comprehensive evaluation of
our models.

\subsection{Transfer-Features} \label{subsection:resultTransferFeature}

\begin{figure}[htbp]
    \includegraphics[width=\linewidth]{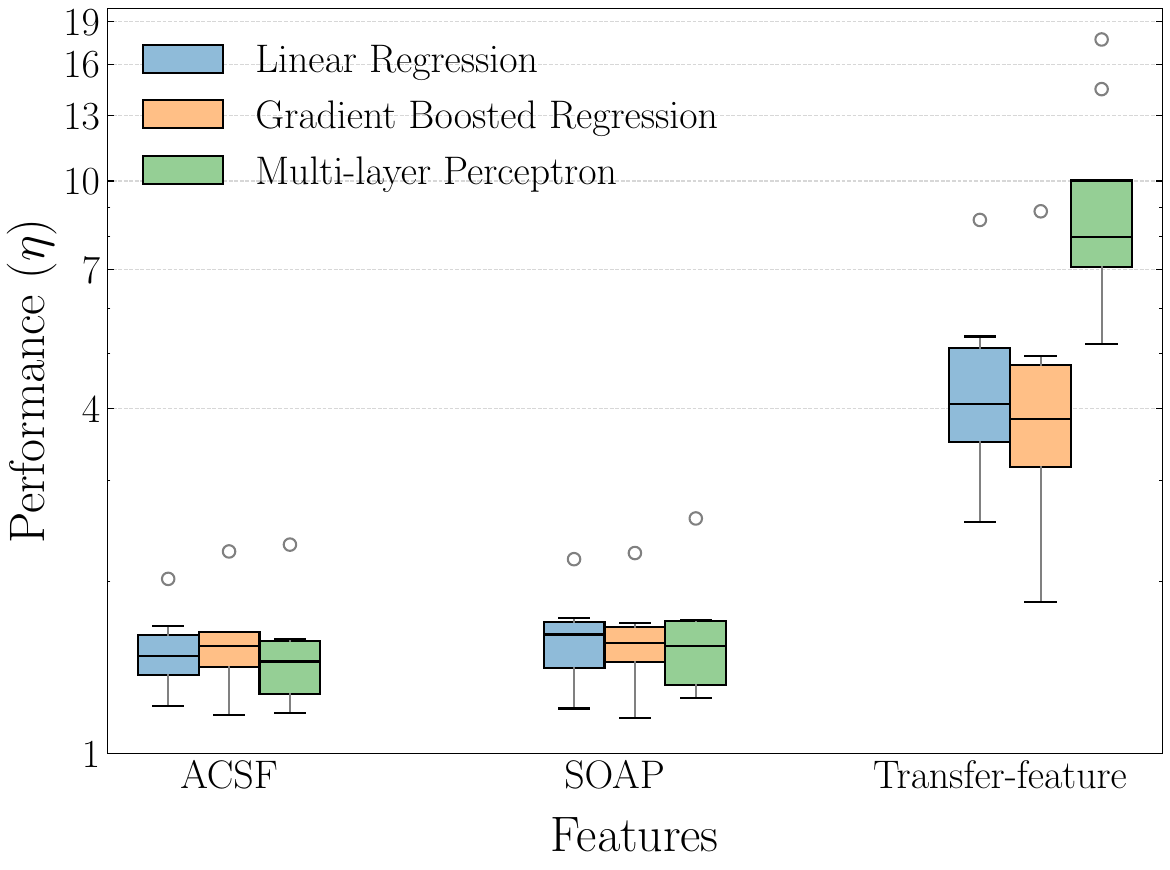}
  \caption{Performance comparison of ML models using different feature sets for
    XAS prediction. The box plots show the performance ($\eta$) of three ML
    models of varied complexity—Linear Regression, Gradient Boosted Regression, and
    Multi-layer Perceptron—trained on three types of features: ACSF, SOAP,
    and transfer-features derived from M3GNet's latent subspace (as used in
    XASModels). The results of the multi-layer perceptron correspond to the ExpertXAS model.}
  \label{fig:featurePerformance}
\end{figure}

Sec.~\ref{subsection:analysisfeatureTransfer} presents qualitative evidence 
supporting our feature transfer hypothesis, which posits that transfer-features 
are effective for XAS prediction. To further examine this hypothesis, we now provide 
quantitative results. Additionally, we compare our featurization approach to 
commonly used methods in the field, specifically the Atom-Centered Symmetry 
Function (ACSF)~\cite{rankine2020deep} and Smooth Overlap of Atomic Positions 
(SOAP)~\cite{rankine2022accurate}.

ACSF and SOAP are established methodologies for generating consistent,
symmetry-invariant representations of local atomic
environments~\cite{behler2011atom,bartok2013representing}. 
ACSF uses tailored radial and
angular functions, while SOAP expands local atomic density using spherical
harmonics. Both methods generate descriptors that preserve essential local
chemical information and ensure rotational and translational invariance. This
allows ML models to learn structure-property relationships without
rediscovering fundamental principles (e.g., symmetry), making them suitable for
predicting properties governed by local atomic environments, such as XAS
spectra~\cite{ghose2023uncertainty}.

Even though both ACSF and SOAP aim to provide descriptors of fixed length, the
resulting representations can get very large and redundant. To overcome this,
we applied PCA to reduce the dimension of the feature vector while
capturing 99\% of the explained variance. This process significantly reduced
the high-dimensional ACSF and SOAP representations (often several thousand
dimensions) to below 100 dimensions.

Here we compare the effectiveness of three different feature sets: ACSF, SOAP,
and M3GNet transfer-features. To avoid potential bias from a particular choice
of the model, we trained three classes of ML models of varied complexity:
a Linear Regression model, Gradient Boosted Regressor, and Multi-layer Perceptron.
For the MLP-based models, we employed the same XAS block architecture, as
described in Sec.~\ref{subsubsection:xasBlock}, thereby making the MLP model trained
on transfer-features correspond to the ExpertXAS model.

The results in Fig.~\ref{fig:featurePerformance} show that ML
models trained on transfer-features to predict XAS consistently yield $\eta$
values much greater than one, across model complexities thereby supporting our hypothesis that transfer-features are effective for
XAS prediction. Furthermore, the performance of the transfer-feature-based
models is significantly better than those trained on ACSF and SOAP across all
model complexities. We will further discuss the implications of these results
in Sec.~\ref{subsection:discussTransferFeature}.

\subsection{Performance Overview} \label{subsection:perfOverview}

\begin{figure}[htbp]
  \includegraphics[width=\linewidth]{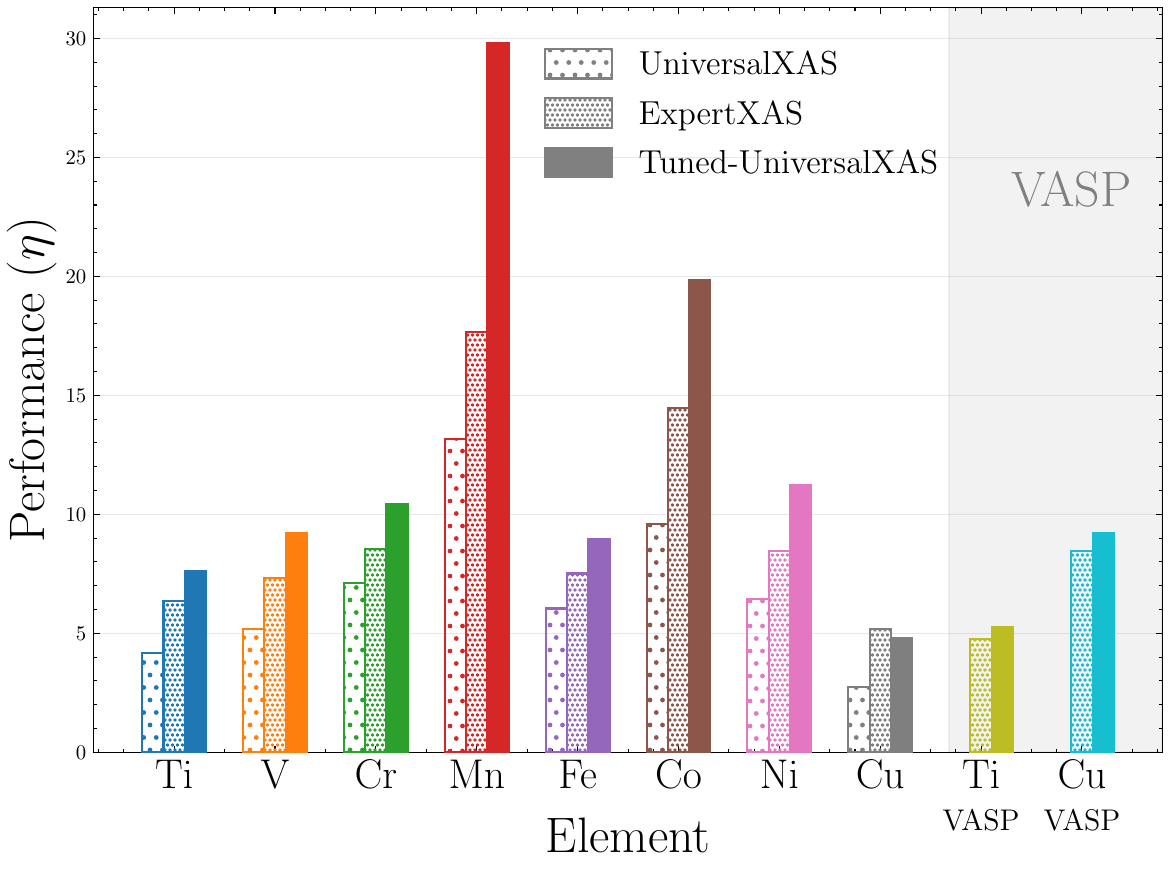}
  \caption{XASModels performances across different elements. The bar chart
    shows the performance ($\eta$) of XASModels: 10 ExpertXAS models, 1
    UniversalXAS model, and 10 Tuned-UniversalXAS models, evaluated across
    various absorbing elements using both FEFF and VASP spectra. The VASP
    models are highlighted with shaded regions. The performance of the
    Universal model is partitioned for each element, as described in Sec.~\ref{subsection:universalXASPerformance}, enabling direct comparison with
    other models. See Table~\ref{tab:model-metrics} for performance metrics.
  }
  \label{fig:performance}
\end{figure}

We now present quantitative results to showcase the general trends in
performance of 21 instances of XASModels. We employ the $\eta$ metric,
visualized in Fig.~\ref{fig:performance}, to evaluate model efficacy. Table~\ref{tab:model-metrics} delineates specific performance metrics and offers
revealing comparative metrics of ExpertXAS and Tuned-UniversalXAS models. Our
investigation expands upon earlier qualitative findings regarding transfer
learning and the potential universality of XAS prediction models, as discussed
in Secs.~\ref{subsection:analysisfeatureTransfer} and
\ref{subsection:analysisHintsOfUniversality}.


\begin{table*}[htbp]
  \caption{Performance metric of ExpertXAS ($\eta_\text{E}$), UniversalXAS ($\eta_\text{U}$), and
    Tuned-UniversalXAS ($\eta_\text{T}$) models. The best performance for each element is
    highlighted in bold. The improvement column shows the percentage increase in
    performance of the Tuned-UniversalXAS model over the ExpertXAS model. The
    Win Rate shows improvement across spectra defined by Eq.~\ref{eq:win_s} and energy grid defined by Eq.~\ref{eq:win_e}.
  }
  \label{tab:model-metrics}
\begin{ruledtabular}
\begin{tabular}{lccccccc}
\toprule
    & \multicolumn{3}{c}{$\eta$} & & \multicolumn{2}{c}{Win Rate (\%)} \\
\cmidrule(lr){2-4} \cmidrule(lr){6-7}
    Element & ExpertXAS & UniversalXAS & Tuned-UniversalXAS & Improvement (\%) & Energy & Spectra \\
    & $\eta_\text{E}$ & $\eta_\text{U}$ & $\eta_\text{T}$ & $(\eta_\text{T}-\eta_\text{E})/\eta_\text{E}$ & $w_\text{e}$ & $w_\text{s}$ \\
\midrule
     Ti      & 6.35  & 4.19 & \textbf{7.63} & 20.15 & 94.33 & 65.21 \\
    V       & 7.30  & 5.19 & \textbf{9.22} & 26.17 & 100.00 & 63.98 \\
    Cr      & 8.54  & 7.13 & \textbf{10.44} & 22.32 & 97.87 & 69.18 \\
    Mn      & 17.66 & 13.15 & \textbf{29.81} & 68.77 & 100.00 & 71.19 \\
    Fe      & 7.51  & 6.04 & \textbf{8.98} & 19.56 & 97.87 & 62.66 \\
    Co      & 14.47 & 9.58 & \textbf{19.83} & 37.07 & 100.00 & 67.78 \\
    Ni      & 8.45  & 6.43 & \textbf{11.21} & 32.71 & 99.29 & 65.74 \\
    Cu      & \textbf{5.19} & 2.75 & 4.81 & -7.42 & 75.89 & 53.37 \\
    Ti VASP & 4.75  & N/A & \textbf{5.27} & 10.82 & 74.47 & 63.13 \\
    Cu VASP & 8.46  & N/A & \textbf{9.21} & 8.92 & 95.74 & 66.04 \\
\bottomrule
\end{tabular}
\end{ruledtabular}
\end{table*}

\subsubsection{ExpertXAS Performance}

ExpertXAS models consistently outperform the baseline across all elements
studied, demonstrating their effectiveness in predicting XAS spectra. As shown in Table~\ref{tab:model-metrics} and Fig.~\ref{fig:performance}, the
model's predictive power varies widely by element, with $\eta$ ranging from 4.75
to 17.66, indicating the accuracy's strong dependence on element type.

For FEFF spectra, Mn and Co show the best performance, with highest $\eta$ reaching up
to 17.66, while Ti and Cu exhibit lower performance, with lowest $\eta$ at 5.19.
Notably, even for Ti and Cu, the performance is more than five times of the
baseline model. The performance trends differ for VASP spectra. Similar to FEFF
models, Ti has an $\eta$ value of 4.75, being the lowest among all models.
However, the performance of Cu VASP is much higher than its FEFF counterpart,
reaching the $\eta$ value of 8.46.

\begin{figure} [htbp]
  \includegraphics[width=\linewidth]{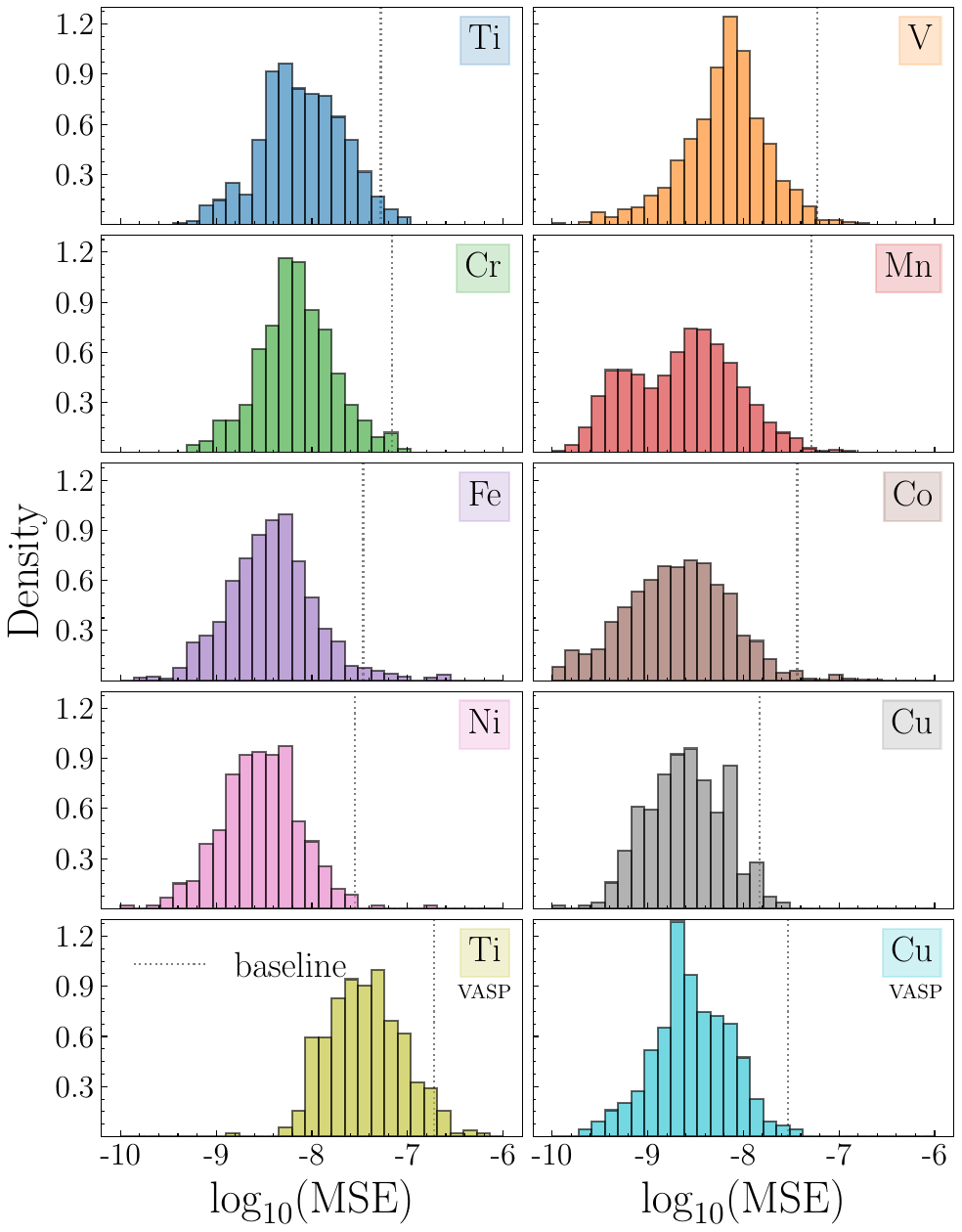}
  \caption{The density distribution of spectral prediction errors (MSE) for
    ExpertXAS models across different elements. The gray dashed line represents
    the baseline MSE ($\xi_\text{baseline}$) defined in Eq~\ref{eq:eta}.
  }
  \label{fig:mseHistExpertXAS}
\end{figure}

The MSE distribution of ExpertXAS models shown in Fig.~\ref{fig:mseHistExpertXAS} demonstrates consistent patterns across different elements when examined on a logarithmic scale (log10(MSE)). These errors exhibit bell-shaped distributions centered between -9 and -7, except for Mn FEFF, which shows a bimodal distribution. Small errors are common, while large errors become exponentially rare. While spanning approximately three orders of magnitude, these distributions consistently show lower errors than their respective baselines. The similarity in distribution shapes and their consistent positioning relative to the baselines across the transition metal series suggests robust model behavior.

\subsubsection{UniversalXAS Performance} \label{subsection:universalXASPerformance}

To facilitate the direct comparison of the UniversalXAS model with other
models, we performed an element-wise breakdown of the UniversalXAS model
performance. We first partitioned the test set into element-specific subsets
and calculated the median MSE for each element separately. These
element-specific values were then normalized by the same $\xi_\text{baseline}$
used for other model variants, derived from the mean spectrum of the entire
training set. This approach yields element-specific $\eta$ values for the
UniversalXAS model, enabling direct comparison with other models despite the
UniversalXAS model being trained on multi-element data with varying spectral
scales.

The UniversalXAS model demonstrates $\eta$ values ranging from 2.75 to 13.15
across different elements as shown in Fig.~\ref{fig:performance} and 
Table~\ref{tab:model-metrics}. While these values are generally lower than those of
the corresponding ExpertXAS models, they consistently exceed the baseline
performance for all elements. Notably, the element-wise performance trend of
the UniversalXAS model closely mirrors that of the ExpertXAS models,
maintaining a similar ranking order but with lower magnitudes.

\subsubsection{Tuned-UniversalXAS Performance}

Tuned-UniversalXAS models outperform both ExpertXAS and UniversalXAS models
across all but one case (Cu FEFF), with varying degrees of improvement.
The smallest improvement is observed for Cu VASP, with a 8.92\% increase in
$\eta$, while the largest improvement is seen for Mn FEFF, with a substantial
68.77\% increase. 
For the exceptional case
of Cu FEFF, the Tuned-UniversalXAS model performs slightly worse than the
ExpertXAS model, with a decrease of 7.42\% in $\eta$.


\begin{figure}[htbp]
  \includegraphics[width=\linewidth]{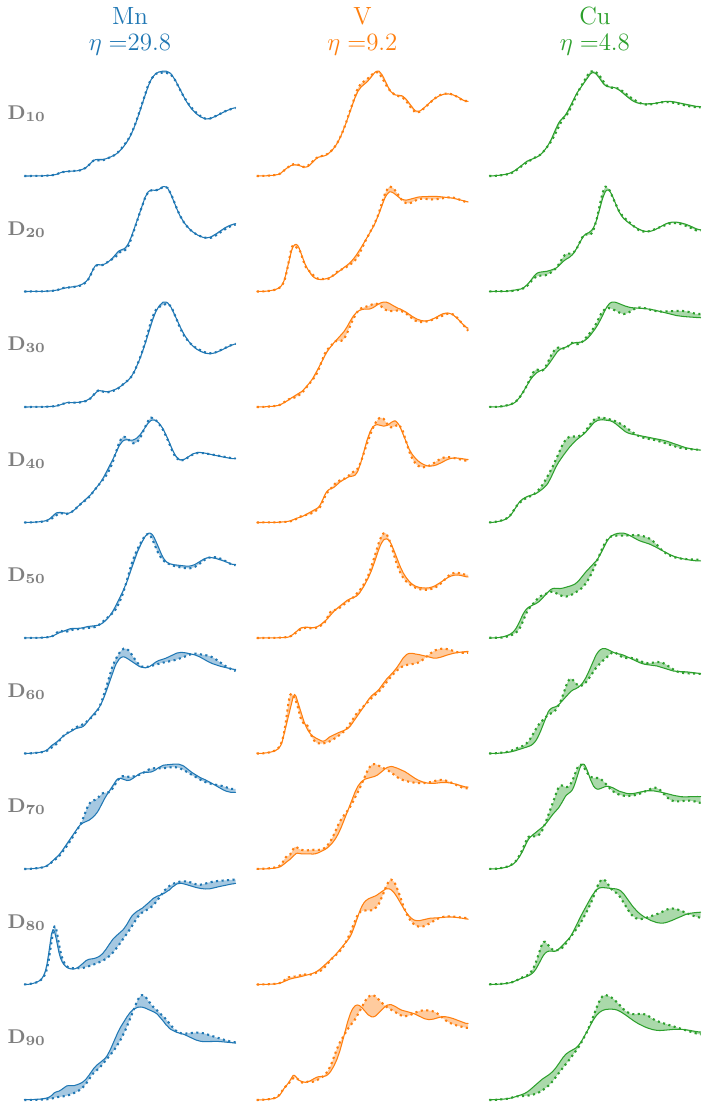}
  \caption{
    Spectra from deciles \( D_i \) sorted by the MSE-per-spectra of the best
    model's (Tuned-Universal) prediction. Solid lines represent simulated
    spectra (ground truth), dashed lines represent predicted spectra, and shaded areas
    highlight the differences. Performance of Mn, V and Cu corresponds to
    the maximum, median, and minimum \(\eta_i\) across elements.
  }
  \label{fig:examples}
\end{figure}

\subsection{Example Predictions}
We now provide specific examples to demonstrate the predictive power of
XASModels. Fig.~\ref{fig:examples} shows FEFF spectral predictions for Mn, V,
and Cu, corresponding to the highest, median, and lowest performing models,
respectively, within the Tuned-UniversalXAS variant. Predictions are
categorized into deciles based on MSE per spectrum, with $D_{i}$ representing
the $i$-th percentile of the MSE distribution. We also provide comprehensive
view of example predictions from nine deciles for each of the ten
element/simulation pairs, totaling 252 spectra in Supplementary Figs.~S1-S3.

As shown in Fig.~\ref{fig:examples}, in the best performing XASModel of Mn,
the predicted spectra closely align with the simulated spectra across all deciles. The shaded areas, representing differences
between predicted and simulated spectra, are minimal even in the lower deciles.
For the representative median performance of V, predictions show
excellent agreement in the higher deciles ($D_{10}$ -- $D_{50}$), with slightly
larger deviations in the lower deciles, particularly in the main-edge and post-edge regions.
For the case of  Cu, which represents the model with lowest performance, predictions
still accurately capture the main edge and overall spectral shape in the higher
deciles, with more noticeable differences in the fine details of the lower
deciles.

Overall key spectral features are consistently reproduced across all three
representative models. Even in the lowest deciles for the lowest-performing
models, the essential characteristics of the XAS spectra (e.g., the overall spectral shape) are captured,
demonstrating the robustness of our models across a wide range of performance
levels.

Given the ultimate goal of ML models as an accurate and efficient tool for analyzing experimental XANES spectra, it is important to understand how accurately ML predictions can reproduce experiment. However, we note that the OmniXAS framework developed here is designed to serve as a surrogate for XANES \textit{simulations}, and as such, the models are not trained to accurately reproduce experimental data to any higher accuracy than FEFF or VASP. Since the quantitative assessment of OmniXAS against the ground truth of the simulations has been rigorously established, a systematic benchmark of simulation (e.g., FEFF and VASP) against experiment is warranted for future studies.

\subsection{Knowledge Transfer} \label{subsection:knowledgeTransfer}
Comparison of models based on the performance metric $\eta$ in Sec.~\ref{subsection:perfOverview} indicates the superiority of Tuned-UniversalXAS
models over ExpertXAS models. In this section, we dissect the general trend of
performance improvement and delve deeper into the nature of these improvements,
analyzing how and where they occur. This analysis is motivated to gain insights
into the extent and impact of knowledge transfer from the UniversalXAS model to
the Tuned-UniversalXAS model.

\subsubsection{Improvement Distribution} \label{subsubsection:transferStatistics}
To evaluate the performance difference between Tuned-UniversalXAS and ExpertXAS models, we compare their predictions on a common set of spectra. Our analysis focuses on the residual differences at each prediction point, which vary with respect to both the individual spectrum and energy. We define our notation as follows:
\begin{align*}
\mu_j^{(i)} &: \text{XAS ground truth (simulation)}, \\
\hat{\mu}_j^{(i,\text{E})} &: \text{ExpertXAS prediction}, \\
\hat{\mu}_j^{(i,\text{T})} &: \text{Tuned-UniversalXAS prediction},
\end{align*}

\noindent where $i$ indexes the spectrum in the test set and $j$ represents the energy grid points within each spectrum. We introduce the following metric of residual difference ($\Delta_j^{(i)}$) to quantify the relative performance of the two models:
\begin{equation}
\Delta_j^{(i)} \equiv |\hat{\mu}_j^{(i,\text{E})}-\mu_j^{(i)}| - |\hat{\mu}_j^{(i,\text{T})}-\mu_j^{(i)}|.
\label{eq:residual_difference}
\end{equation}

\begin{figure} [hbtp]
  \includegraphics[width=\linewidth]{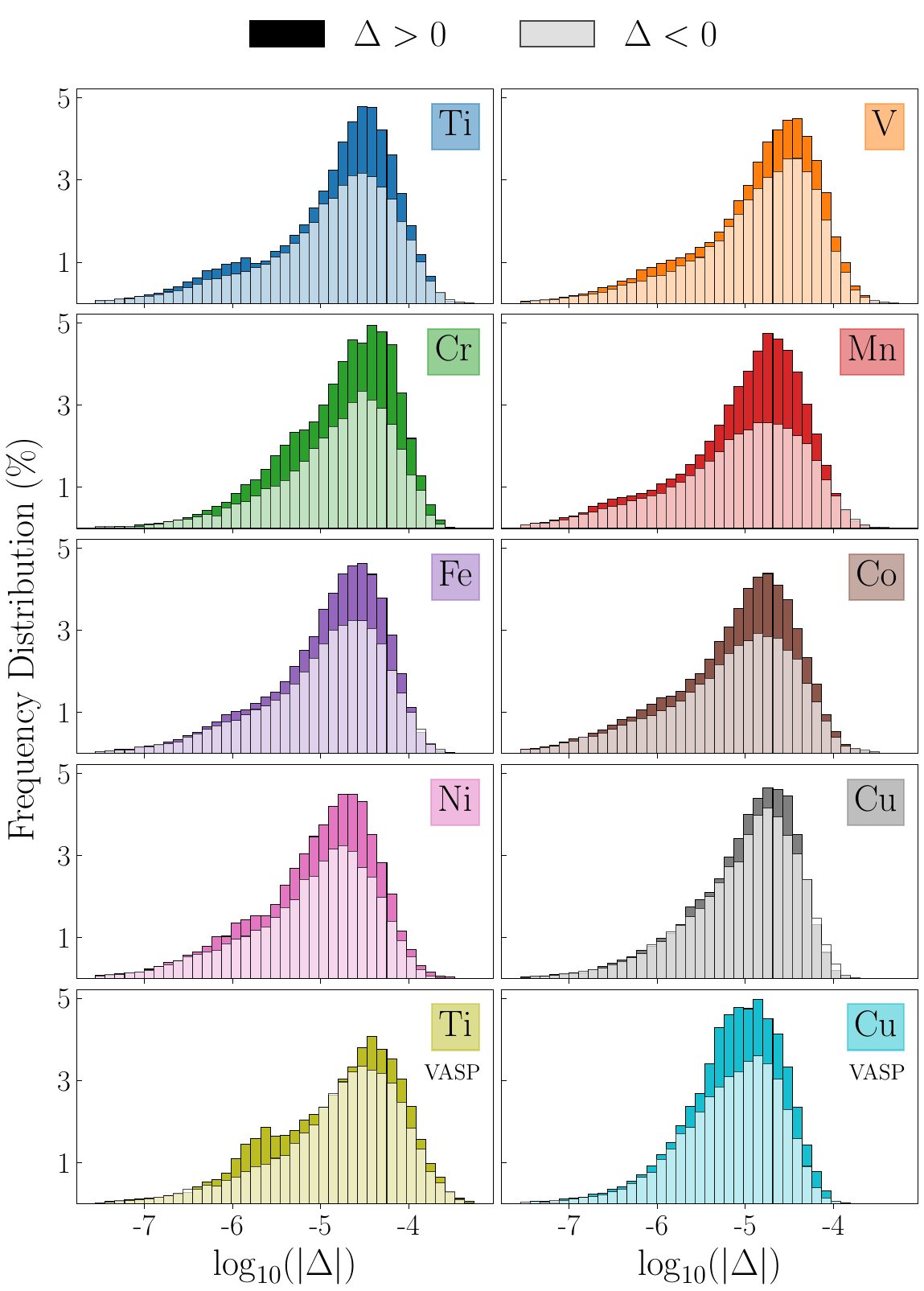}
  \caption{Distribution of residual differences 
  $ \Delta = \{ \Delta_j^{(i)} \mid i = 1, 2, \ldots, N; \, j = 1, 2, \ldots, M \}$
  between Tuned-UniversalXAS and ExpertXAS predictions (see Eq.~\ref{eq:residual_difference}). 
  For easy comparison between improvements and degradations, the data is partitioned by sign of $\Delta_j^{(i)}$ and plotted against its absolute value in logarithmic scale. Darker bars ($\Delta_j^{(i)} > 0$) indicate predictions where Tuned-UniversalXAS outperforms ExpertXAS, while lighter bars ($\Delta_j^{(i)}<0$) indicate the opposite. Bar heights represent the percentage of the data points, with the total number of data points given by the product of the number of energy grids and the spectra count. 
    }
  \label{fig:residualDifference}
\end{figure}

This metric provides a point-by-point comparison of model performance. Cases where $\Delta_j^{(i)} > 0$ indicate superior performance by the Tuned-UniversalXAS model, while $\Delta_j^{(i)} < 0$ favors the ExpertXAS model.
Fig.~\ref{fig:residualDifference} illustrates the distribution of all residual differences partitioned by its sign.
We observe dominant instances of performance improvements (darker bars) compared to performance degradations (lighter bars) for most elements. 
Notably, for Cu, where the difference in $\eta$ value is small, we see a more balanced distribution between improvements and degradations. These observations aligns with earlier observations in Sec.~\ref{subsection:perfOverview} using the $\eta$ metric.


\subsubsection{Win Rates} \label{subsubsection:win_rates}
To further quantify the  performance differences between the Tuned-UniversalXAS and ExpertXAS models, we introduce two complementary ``\emph{win rate}" metrics. These metrics are based on the residual difference, $\Delta_j^{(i)}$, defined in Eq.~\ref{eq:residual_difference}. Both metrics employ the standard indicator function, denoted as $\mathbb{1}(x)$, which is defined as
\begin{equation}
\mathbb{1}(x) = \begin{cases} 
1 & \text{if } x \text{ is true}, \\
0 & \text{if } x \text{ is false}.
\end{cases}
\end{equation}
Our first metric, the \emph{win rate across spectra} ($w_\text{s}$), provides 
a assessment of performance differences among model in terms of spectra predictions,
\begin{equation}
w_\text{s} = \frac{1}{N} \sum_{i=1}^{N} \mathbb{1}\left(\sum_{j=1}^{M} \Delta_j^{(i)} > 0\right).
\label{eq:win_s}
\end{equation}
This metric calculates the proportion of spectra where the Tuned-UniversalXAS model outperforms the ExpertXAS model when considering the entire energy range.

To complement the broader view presented by $w_\text{s}$, we introduce the 
\emph{win rate across energy} ($w_\text{e}$), which offers a more granular view based on energy grid,
\begin{equation}
w_\text{e} = \frac{1}{M} \sum_{j=1}^{M}\mathbb{1}\left( \frac{1}{N} \sum_{i=1}^{N} \Delta_j^{(i)} > 0\right).
\label{eq:win_e}
\end{equation}
This metric represents the fraction of energy grids where the Tuned-UniversalXAS model outperforms the ExpertXAS model, when averaged over all spectra.

Both $w_\text{s}$ and $w_\text{e}$ provide valuable insights into model performance. A $w_\text{s}$ value exceeding 0.5 indicates that the Tuned-UniversalXAS model performs better on the majority of spectra. Similarly, a $w_\text{e}$ value above 0.5 suggests superior performance of the Tuned-UniversalXAS model at the majority of energy grid. 

As shown in Table~\ref{tab:model-metrics}, all $w_\text{s}$ are larger than 0.5 ranging from 53.37\% in Cu FEFF to 71.19\% in Mn FEFF, suggesting a consistent spectrum-wise performance improvement of the Tuned-UniversalXAS model over the ExpertXAS model. The overall trend in $w_\text{s}$ is in line with $(\eta_\text{T}-\eta_\text{E})/\eta_\text{E}$. It is worth noting that in Cu FEFF, although $(\eta_\text{T}-\eta_\text{E})/\eta_\text{E}$ yields a negative value of -7.42 based on the median MSE, $w_\text{s}=53.37\%$ indicates a slight improvement based on more fine-grained spectrum-wise metric. The advantage of the Tuned-UniversalXAS model is further supported by $w_\text{e}$, which shows a more significant performance improvement in the energy-wise metric with values varying from 74.47\% in Ti VASP to 100\% in V, Mn and Co FEFF. Similar to $w_\text{s}$, a $w_\text{e}$ of 75.89\% in Cu FEFF suggests a significant energy-wise performance improvement of the Tuned-UniversalXAS model over the ExpertXAS model.

\subsubsection{Energy-resolved Improvements} \label{subsubsection:transferLocalization}
\begin{figure}[htbp]
  \centering
  \includegraphics[width=0.98\linewidth]{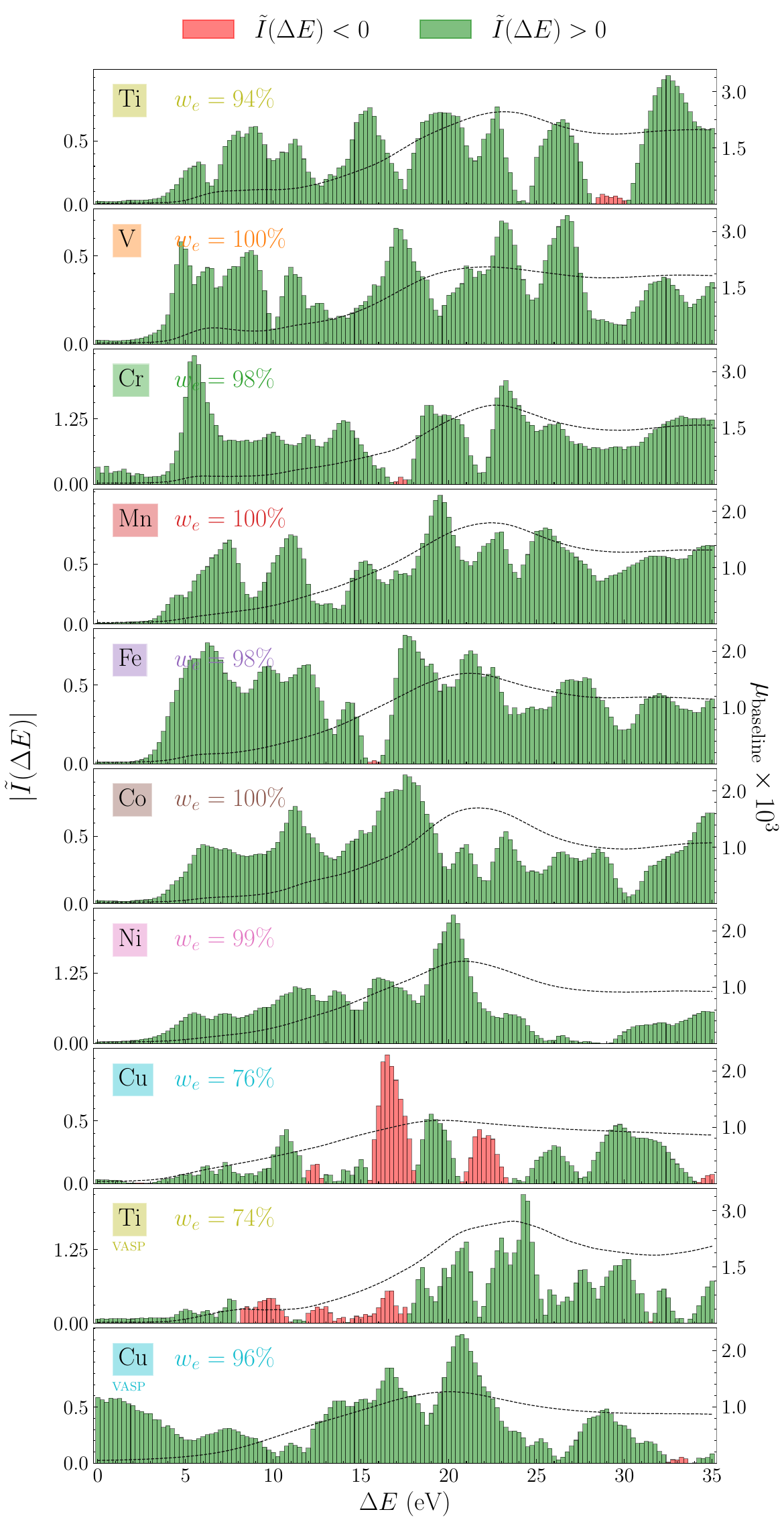}
  \caption{
    Energy-resolved performance improvement
    $\tilde{I}(\Delta E)$ when using the Tuned-UniversalXAS model over the
    ExpertXAS model. The regions of improvement in Tuned-UniversalXAS
    predictions are highlighted in green, while the regions of degradation are
    highlighted in red. The spectrum used for normalization 
    ($\mu_{\text{baseline}}$) is overlaid
    as a dotted line corresponding to the right y axis.
  }
  \label{fig:performanceAcrossEnergy}
\end{figure}
To further quantify the energy-wise performance metric associated with $w_\text{e}$, we compare the residual differences between prediction of ExpertXAS and Tuned-UniversalXAS models averaged over all spectra to examine where in the energy region those improvement occurs. For this purpose, we define an energy-resolved improvement metric $I(\Delta E_j)$ as follows:

\begin{equation}
    I(\Delta E_j) \equiv \frac{1}{N} \sum_{i=1}^N \Delta_{j}^{(i)},
\end{equation}
where $\Delta_{j}^{(i)}$ is the residual difference defined in Eq.~\ref{eq:residual_difference}.

To put the significance of varying levels of improvement among spectra of different scales from different elements into perspective, we normalize $I(\Delta E)$ using the baseline model introduced in Sec.~\ref{subsection:resultPerformanceMetric} as follows:

\begin{align}
    \bar{\mu}_{\text{baseline}} &= \frac{1}{M}\sum_{j=1}^{M} \left[\mu_{\text{baseline}}\right]_j, \\
    \tilde{I}(\Delta E) &= \frac{I(\Delta E)}{\bar{\mu}_{\text{baseline}}} \times 100.
\end{align}
Positive values of $\tilde{I}(\Delta E)$ indicate better performance by the Tuned-UniversalXAS model, while negative values indicate better performance by the ExpertXAS model.


Fig.~\ref{fig:performanceAcrossEnergy} displays $|\tilde{I}(\Delta E)|$ as bar plots for each element, partitioned by the sign of $\tilde{I}(\Delta E)$. To provide context on the practical significance of the improvement, we overlay $\mu_{\text{baseline}}$ (dotted line), which serves as a reference line to cross-relate the performance improvements to regions of practical significance. For all cases, except Cu FEFF and Ti VASP, the green bars dominate $|\tilde{I}(\Delta E)|$, showcasing that
the Tuned-UniversalXAS model consistently outperforms the ExpertXAS model across wide rage of energy grid. The result also aligns with earlier observations based on $w_\text{e}$, where Cu FEFF and Ti VASP have the lowest scores. Furthermore, cross-comparison with the average spectrum $\mu_{\text{baseline}}$ reveals that the improvements are more pronounced around regions of the main edge and post-edge that are of high importance in XANES.

\subsection{Knowledge Transfer Across Chemical Environments} 
\label{subsection:descriptorTransfer}

We now examine how the knowledge transfer varies across different local chemical environments. This analysis provides insight into which structural motifs benefit most from the general spectral trends captured by our cascaded transfer learning approach.

To analyze the variation in knowledge transfer effectiveness across different 
chemical environments, we examine model performance for specific local descriptors (OS, CN, and OCN) 
previously discussed in Sec.~\ref{subsection:analysisfeatureTransfer}. For each element, 
we partition both training and test datasets into subgroups based on descriptor values. 
The environments with fewer than 10 samples are excluded from our analysis to ensure 
statistical reliability.

For each specific chemical environment characterized by descriptor value $x$ 
(e.g., coordination number $\text{CN} = 6$), we extend our performance metric 
approach from Sec.~\ref{subsection:resultPerformanceMetric} by defining a 
subgroup baseline spectrum. Similar to the baseline spectrum used in the 
original $\eta$ metric (Eq.~\ref{eq:eta}), this subgroup baseline is calculated 
as the mean of all training spectra within that specific chemical environment:

\begin{equation}
  [\mu_\mathrm{baseline}]_j^{(x)} = \frac{1}{N_{\text{train}}^{(x)}} \sum_{i=1}^{N_{\text{train}}^{(x)}} 
  \mu_j^{(i, x)},
\end{equation}
where $N_{\text{train}}^{(x)}$ is the number of training samples with descriptor value $x$.

For a model $M$ (either ExpertXAS or Tuned-UniversalXAS), we compute the median 
MSE over test samples in subgroup $x$:

\begin{equation}
  \xi^{(M, x)} = \mathrm{Median}(\{J^{(i,x)}\}_{i=1}^{N_{\text{test}}^{(x)}}),
\end{equation}
where $J^{(i,x)}$ is the MSE of the $i$-th spectrum in the test set having descriptor value $x$ out of 
a total of $N_{\text{test}}^{(x)}$ spectra.

We then define the environment-specific performance metric:
\begin{equation}
  \eta^{(M, x)} = \frac{\xi_\mathrm{baseline}^{(x)}}{\xi^{(M, x)}},
\end{equation}
where $\xi_\mathrm{baseline}^{(x)}$ is the median MSE of the subgroup baseline model.

To quantify the improvement of Tuned-UniversalXAS ($T$) over ExpertXAS ($E$) 
for a specific chemical environment characterized by value $x$, we calculate
\begin{equation}
  \tilde{\eta}^{(x)} = \frac{\eta^{(T, x)} - \eta^{(E, x)}}{\eta^{(E, x)}} 
  \times 100.
\end{equation}
A positive $\tilde{\eta}^{(x)}$ indicates that Tuned-UniversalXAS outperforms 
ExpertXAS for the chemical environment characterized by $x$, with larger values 
representing greater improvement. Conversely, negative values indicate performance 
degradation where the ExpertXAS model performs better than the Tuned-UniversalXAS model. 

\begin{figure}[htbp]
  \centering
  \includegraphics[width=\linewidth]{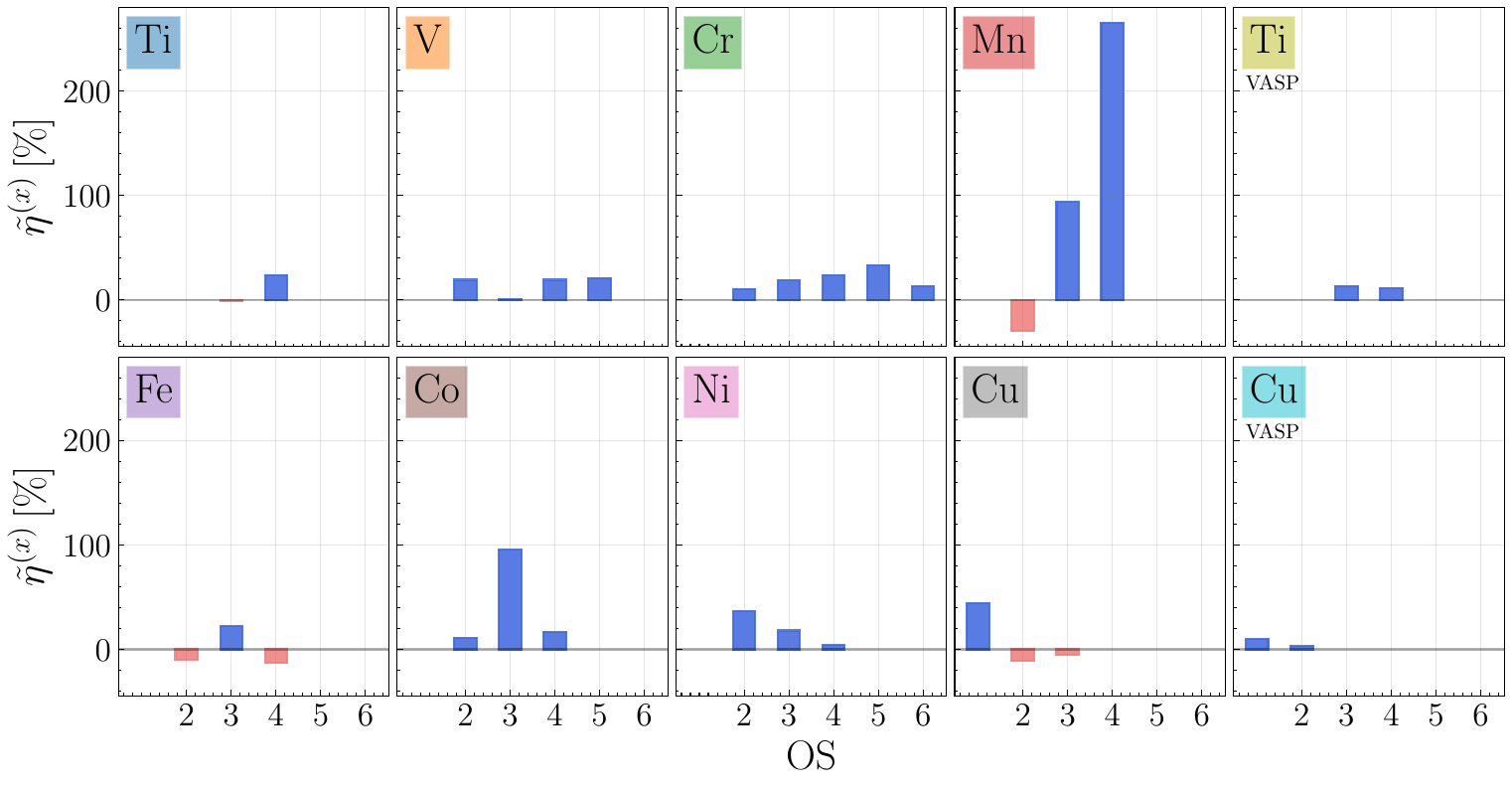}
  \caption{Performance improvement across oxidation states (OS). Positive values (blue
bars) indicate Tuned-UniversalXAS outperforms ExpertXAS; negative values (red bars) indicate the opposite.}
  \label{fig:improvement-OS}
\end{figure}

\begin{figure}[htbp]
  \centering
  \includegraphics[width=\linewidth]{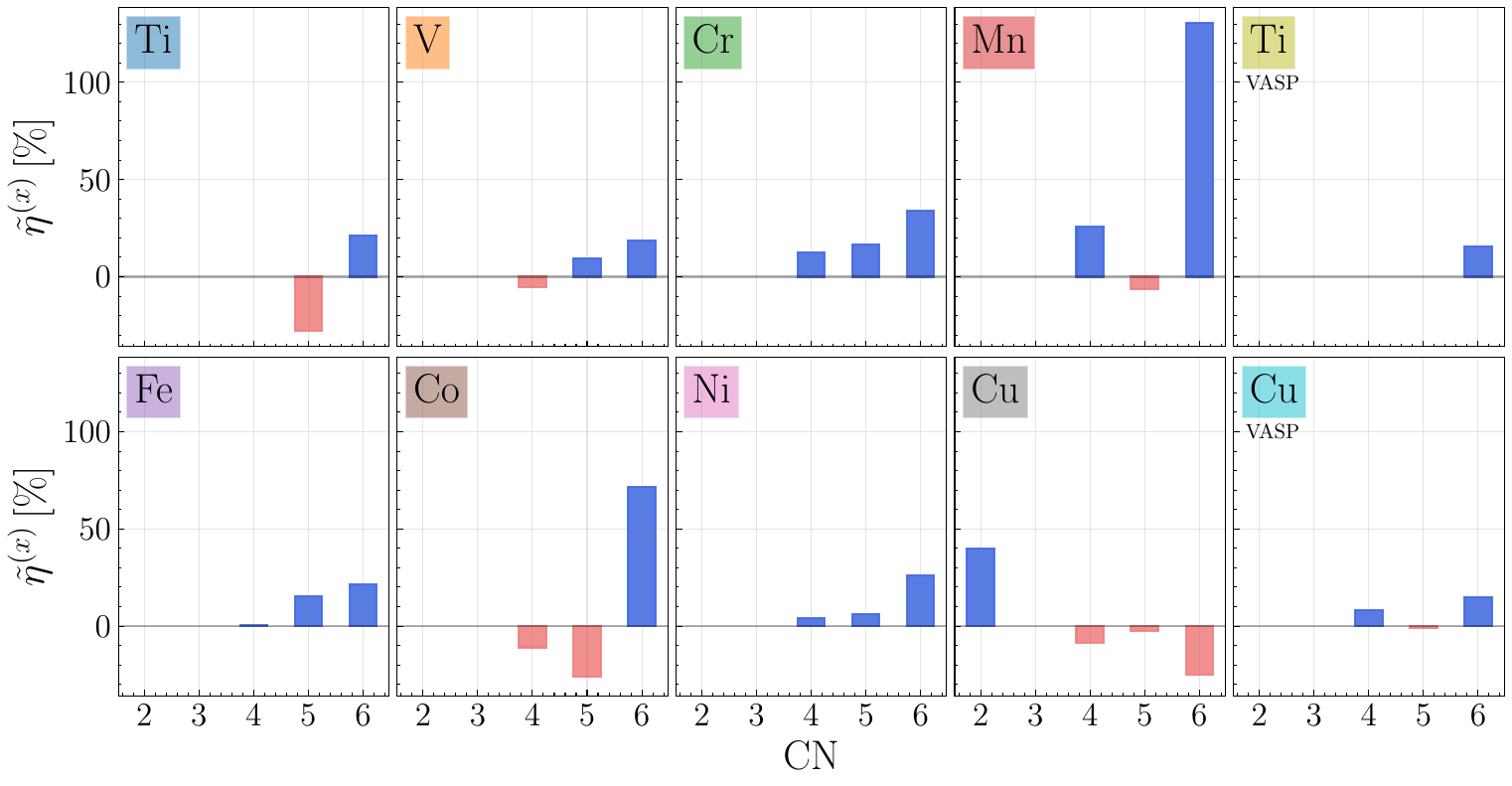}
  \caption{Performance improvement across coordination numbers (CN). Same format as Fig.~\ref{fig:improvement-OS}.}
  \label{fig:improvement-CN}
\end{figure}

\begin{figure}[htb]
  \centering
  \includegraphics[width=\linewidth]{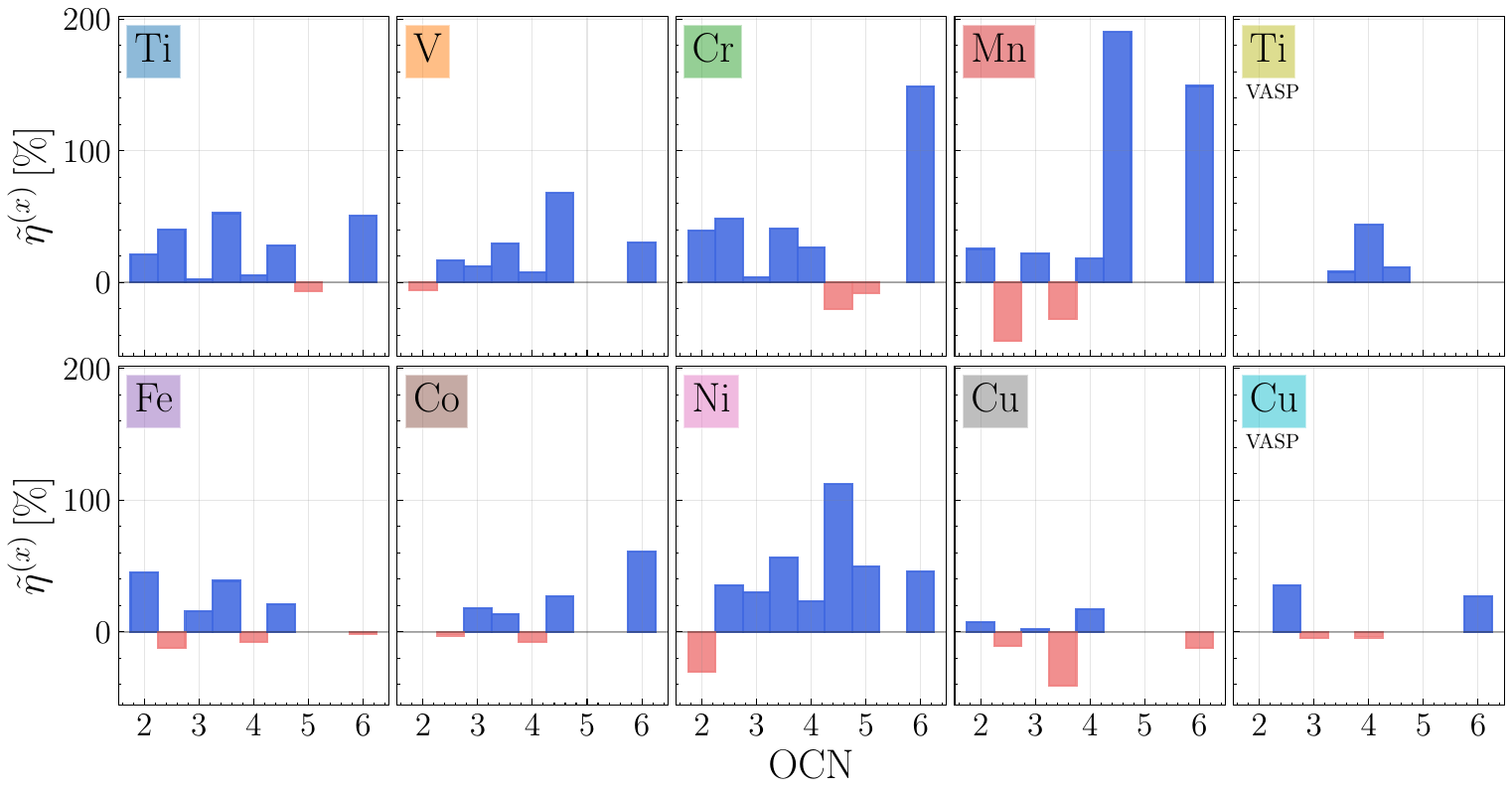}
  \caption{Performance improvement across oxygen coordination numbers (OCN). Same format as Fig.~\ref{fig:improvement-OS}.}
  \label{fig:improvement-OCN}
\end{figure}

Figs.~\ref{fig:improvement-OS}~-~\ref{fig:improvement-OCN} display $\tilde{\eta}^{(x)}$ across OS, CN and OCN. We can see that the majority of the results in OS (Fig.~\ref{fig:improvement-OS}) show an improvements, with the magnitude varying dramatically across elements and its chemical environments. Mn with OS=4 shows exceptional improvement exceeding 250\%.  Very high improvement near 100\% are also found in Mn and Co with OS=3. In other systems, most of the improvements are below 20\%. Only a few systems show small negative $\tilde{\eta}^{x}$, e.g., in Mn (OS=2), Fe (OS=2, 4) and Cu (OS=2, 3), where transfer-learning from the universalXAS model slightly adversely affected the model performance. Similarly, CN results in Fig.~\ref{fig:improvement-CN} overall show a consistent improvement, with the max $\tilde{\eta}^{x}$ of about 130\% in Mn with CN=6. According to the UMAP analysis in Fig~\ref{fig:umapFEFF}g (blue color), CN=6 data form a very diffusive distribution nearly spreading across the entire spectral dataset. This suggests that knowledge transfer in CN=6 could be very efficient. Indeed, a strong correlation between positive $\tilde{\eta}^{x}$ values with CN=6 is identified for all the elements, except for Cu where its CN=6 UMAP cluster is not as obvious as other elements. In comparison to OS and CN, $\tilde{\eta}^{x}$ of OCN exhibit pronounced improvement more broadly
distributed at almost all OCN values as shown in Fig.~\ref{fig:improvement-OCN}. This trend is likely attributed to the very diffusive nature of the OCN spectral UMAP pattern in most of the OCN values as shown in Fig.~\ref{fig:umapFEFF}h.

\section{Discussion} \label{section:discussion}

In this section, we reflect on the broader significance of our findings. In
Sec.~\ref{subsection:discussTransferFeature}, we consider the impact of
transfer-features for XAS prediction, highlighting their practical advantages
and how they contribute to the improved model performance. In Sec.~\ref{subsection:discussVariation}, we delve into the observed variations in
model performance across elements, discussing the factors that may explain
these differences and their implications for generalization. The discussion
continues with an examination of how the UniversalXAS model bridges the
elemental divide, offering potential for more widespread applications across
different materials in Sec.~\ref{subsection:discussCrossElement}.
Furthermore, we assess how knowledge transfer from UniversalXAS enhances the
performance of ExpertXAS models in Sec.~\ref{subsection:disscussKnowledgeTransfer}. In Sec.~\ref{subsection:discussFidelity}, we address
the model's ability to cross fidelity boundaries, demonstrating the utility of
this approach across different simulation environments. Finally, we highlight the huge computational speed up of XANES spectral prediction 
with OmniXAS as compared to first-principles simulations in Sec.~\ref{subsection:discussComputation}.

\subsection{Transfer-Features: Efficacy and Beyond} \label{subsection:discussTransferFeature}

The efficacy of transfer-features for XAS prediction is remarkably evident across various model complexities, as
demonstrated in Fig.~\ref{fig:featurePerformance}. Training of M3GNet on
inter-atomic potential prediction has evidently created a rich representation of
materials structures that translates exceptionally well to XAS prediction tasks.
While some degree of success
was anticipated given the shared focus on local atomic environments, the extent
of the transfer-features' efficacy is particularly noteworthy.

Transfer-features demonstrate a marked superiority in performance compared to
well-established structural descriptors, such as ACSF and SOAP. Although all these methods
aim to characterize local atomic environments, the distinct edge exhibited by
transfer-features suggests they capture additional, highly relevant
information. The hierarchical architecture of M3GNet's graph neural network
may play a crucial role in this success, enabling a multi-scale representation
that adeptly incorporates both local and intermediate-range interactions.

Beyond their predictive power, transfer-features offer several practical
advantages, over ACSF and SOAP, that make them particularly suited for
large-scale XAS predictions. Particularly:

\begin{itemize}
    \item \emph{Compactness}--- Transfer features exhibit higher information density for XAS
        prediction compared to ACSF and SOAP. When subjected to dimensionality
        reduction techniques like PCA (preserving 99\% of explained variance), ACSF
        and SOAP features undergo a significant reduction—approximately two orders
        of magnitude—from their initial representations in the thousands to below
        100 dimensions (see Supplementary Fig.~S4). In contrast, transfer
        features, originating from a compact 64-dimensional space, experience only
        minimal reduction. This indicates that transfer features contain less
        redundancy and encode XAS-relevant information in a more compact manner.
    \item \emph{Scalability}--- Transfer-features offer computational speed advantages that
        scale well with system size and dataset complexity. The transfer-features
        can be generated for large datasets due to native GPU support. 
        Moreover,  transfer-features are compact
        and do not need to undergo additional dimensionality reduction
        post-processing before practical use, a process that typically involves
        computationally expensive techniques like PCA that scale cubically with
        feature dimension~\cite{abraham2014fast}. 
    \item \emph{Efficiency}--- Even simple linear models exhibit enhanced
        performance when trained with transfer-features as shown in Fig.~\ref{fig:featurePerformance} and Supplementary Table S7. This opens up possibilities
        for more accessible and computationally efficient ML modeling for XAS
        prediction, particularly beneficial for larger-scale studies or
        resource-constrained environments.
\end{itemize}

These practical advantages, combined with their strong predictive performance,
position transfer-features as a promising approach for large-scale XAS
predictions across materials.

\subsection{Variations in Model Performance} \label{subsection:discussVariation}

While all ExpertXAS models performed above baseline, their performance varies
widely across elements, with $\eta$ ranging from 4.75 to 17.66, warranting a
discussion of the underlying causes.

As is typical in machine learning, larger datasets generally lead to better
performance, and we observe somewhat similar trend here. For instance, Mn, with
the largest dataset exhibits the best performance. However, this
relationship is not entirely trivial. For example, Co and V have nearly
identical dataset sizes (10,753 vs. 10,813), yet their performance differs
significantly (14.47 vs. 7.30). This suggests that dataset size alone does not
fully account for the performance variation. We provide Supplementary Fig.~S5
for illustrative purposes to showcase this empirical observation.

Beyond dataset size, other factors such as the complexity of local atomic
environments and variations in spectra can also influence performance.
Similarly, the transferability of local encodings from inter-atomic potential prediction to XAS
prediction can vary across elements, affecting the model's accuracy. This
interplay of model performance with dataset size, spectral complexity, and
feature transfer is non-trivial and intersects with the broader domain
of neural scaling laws~\cite{bahri2021explaining} that is beyond the scope of
this work.

\subsection{Crossing the Elemental Divide} \label{subsection:discussCrossElement}

The UniversalXAS model, trained on all FEFF data simultaneously, operates as a
multi-task model designed to generalize across multiple elements. Although its
performance is generally lower compared to the element-specific ExpertXAS
models—an expected outcome given the broad generalization goal—what is
particularly noteworthy is that UniversalXAS still performs well above the
baseline for all elements. This demonstrates its robustness and effectiveness
as a generalizable XAS prediction framework.

The success of UniversalXAS challenges the traditional element-by-element
approach for XAS analysis, showing that a model can predict spectra for a wide
range of materials without sacrificing significant accuracy. Interestingly, the
performance trends of UniversalXAS closely mirror those of the ExpertXAS
models. The elements where ExpertXAS performed well, such as Mn and Fe, are
also the ones where UniversalXAS achieved higher accuracy. This parallel behavior
suggests that while UniversalXAS captures shared spectral characteristics
across elements, it still retains the ability to identify and adapt to
element-specific features that define each element's XAS spectra.

\subsection{Transfer of Universal Knowledge} \label{subsection:disscussKnowledgeTransfer}

The Tuned-UniversalXAS model demonstrates significant performance improvements
after being fine-tuned for individual elements, which aligns with expectations.
However, the most striking result is its consistent ability to surpass the
performance of element-specific ExpertXAS models (with one exception for Cu
FEFF). This outcome highlights a key finding of this study.
Considering the novelty of this approach, we delve deeper to establish the
significance of this observation, the consistency of the improvements, and the
practical relevance of these gains.

The first point to note regarding the improvement in $\eta$ values is that
while there is a clear performance enhancement, the extent of this improvement
varies across elements—a theme consistent with the broader discussion on model
performance variations. While Fig.~\ref{fig:performance} and Table~\ref{tab:model-metrics} 
clearly show the extent of improvement in $\eta$, a caveat to note is that similar improvements in $\eta$ do not
necessarily come with the same level of confidence due to variations in dataset
sizes. To ensure that these improvements are not artifacts of random variation,
we conducted a paired difference test, which confirmed the statistical
significance of the observed gains (refer to Supplementary Fig.~S6 for more details).

In Sec.~\ref{subsubsection:transferStatistics}, we examined whether the
performance differences based on the $\eta$ metric alone might be concealing
any systematic biases or outliers. Fig.~\ref{fig:residualDifference} compares
the differences between the Tuned-UniversalXAS and ExpertXAS models across the
entire energy grid for all spectra. The distribution of the residual
differences reveals a clear pattern: when considering each prediction point, the
changes in prediction accuracy are predominantly positive, spread across the
energy grid, and not driven by isolated outliers or noise.

The results from Sec.~\ref{subsubsection:transferLocalization} further
elucidate the nature of these improvements, showing that the regions of
improvement within the spectra are consistent with the regions of importance in
XAS (e.g., main edge and post edge). This alignment underscores the practical
relevance of the performance gains, as they are concentrated in the critical
regions needed for accurate XAS analysis. The win rates (percentage of spectra
where the Tuned-UniversalXAS model outperforms the ExpertXAS model) shown in
the the same Fig.~\ref{fig:performanceAcrossEnergy} shed insight into how
these localized gains translate into improved prediction accuracy at the
spectrum level.

\subsection{Crossing the Fidelity Divide} \label{subsection:discussFidelity}

The performance improvement of the Tuned-UniversalXAS model over both VASP
datasets is clear and definitive, showing a 10.82\% improvement for Ti and
8.92\% for Cu. This is further supported by spetrum-wise win rates (63.13\% for Ti
and 66.04\% for Cu), as well as consistent improvements across energy points
and key regions of significance.

This result represents a distinctive benefit in crossing the fidelity divide
between FEFF and VASP simulations. While previous efforts focused on
fine-tuning within a single fidelity (FEFF), the current findings demonstrate
how a cascading transfer learning approach can effectively transfer knowledge
from low-cost FEFF simulations to the more computationally intensive VASP
simulations, offering substantial practical advantages for efficient and scalable predictions.

\subsection{Computational Performance} \label{subsection:discussComputation}

We envision that the proposed OmniXAS framework provides a practical solution to tackle the computational barrier in XAS spectral analysis.
In average, an XASModel takes only a fraction of second (about 0.16 second) on an Apple M2 Max chip to predict a material XANES spectrum, with most of the time spent on featurization (i.e., the M3GNet Block).
On the other hand, performing a XANES spectral simulation using VASP requires high-performance computing facilities and an exemplary calculation on Ti K-edge XANES of anatase TiO$_2$ takes about 11.57 CPU hours on Intel(R) Xeon(R) Gold 6336Y processors. Therefore, deep learning XASModels can lead to a remarkable speed up of over five orders of magnitude in XAS prediction. The huge computational speed up of the OmniXAS framework should be comparable or better than typical graph neural network potential frameworks, because first-principles spectral calculations are more expensive than total energy calculations of the ground state.  Such a highly efficient XAS prediction capability allows high-throughput XAS modeling applied to a broad material space and opens the door for real-time XAS analysis.

\section{Conclusion} \label{section:conclusionandfuture}
In this study, we developed OmniXAS, a universal deep learning framework for predicting
K-edge X-ray absorption spectra of eight 3d transition metals (Ti -- Cu) based on the 3dtm\_xanes\_ml2024 dataset, utilizing
a cascading combination of inductive transfer learning and domain adaptation
techniques. The integration of M3GNet-derived latent features, referred to as
transfer-features, consistently outperformed traditional methods like ACSF and
SOAP, demonstrating clear advantages for large-scale applications.

We demonstrate that the concept of universal XAS model can indeed be established for light transition metals, based on common spectral trends associated with the local chemical descriptors in this class of materials. This is an important initial step towards developing a fully universal ML framework that covers most of the periodic table.
By training a single universal model, we successfully addressed the elemental
divide often encountered in machine learning models for materials science,
establishing a robust foundation for general-purpose XAS predictions across
diverse elements. 

Our non-linear embedding analysis of the latent space further reinforced the
efficacy of this approach. Fine-tuning the UniversalXAS model into
Tuned-UniversalXAS models led to substantial performance improvements across
nearly all elements, with these gains being both quantitatively significant and relevant for practical applications.
Furthermore, the method demonstrated its ability to extend across fidelity
boundaries, bringing higher performance efficiency from limited amount of computationally intensive dataset.
This transfer learning framework is generalizable to enhance deep-learning models that target other properties in materials research.


\section*{Data \& Software Availability}

The software implementation associated with this study is openly available on GitHub (\url{https://github.com/AI-multimodal/OmniXAS})~\cite{omniXAS}. The 3dtm\_xanes\_ml2024 dataset used in the machine learning model development of this work is available in the Zenodo repository (
\url{https://zenodo.org/records/14145457} 
). Input and output of the spectral simulation will be made available through the Materials Cloud platform~\cite{talirz2020materials} (\url{ https://doi.org/10.24435/materialscloud:85-8x}).

\begin{acknowledgements}

This research is supported by the U.S. Department of Energy,
Office of Science, Office Basic Energy Sciences, Award Number FWP
PS-030. This research used theory and computation resources of the
Center for Functional Nanomaterials, which is a U.S. Department of Energy
Office of Science User Facility, and the Scientific Data and Computing
Center, at Brookhaven National Laboratory (BNL) under Contract No. DE-SC0012704. M.~R.~C. acknowledges
BNL Laboratory Directed Research and Development (LDRD) grant no. 24-004. This research also used resources of the National Energy Research Scientific Computing Center (NERSC), a U.S. Department of Energy Office of Science User Facility located at Lawrence Berkeley National Laboratory, operated under Contract No. DE-AC02-05CH11231 using NERSC Awards No. BES-ERCAP-20690, 18006, and 14811. We thank Mark Hybertsen and Gerald T. Seidler for helpful discussions. We would also like to extend our gratitude to Nina Cao and Pavan Ravindra for their valuable feedback on manuscript.

\end{acknowledgements}



%
\end{document}